\documentclass[useAMS,usenatbib]{mnras}
\usepackage{aas_macros}
\usepackage{natbib}
\usepackage{amsfonts}
\usepackage{amsmath}
\usepackage{amssymb}
\usepackage{graphicx}
\usepackage{color}
\usepackage[dvipsnames]{xcolor}
\usepackage{microtype}

\hypersetup{colorlinks=true,allcolors=teal}

\newcommand{\p}{\ensuremath{\partial}}

\newcommand{\Mh}{\ensuremath{h^{-1}M_{\odot}}}

\newcommand{\Mpch}{\ensuremath{h^{-1}{\rm Mpc}}}
\newcommand{\kpch}{\ensuremath{h^{-1}{\rm kpc}}}

\newcommand{\mb}[1]{\ensuremath{\mathbf{#1}}}

\newcommand{\avg}[1]{\ensuremath{\left\langle \,#1\, \right\rangle}}
\newcommand{\e}[1]{\ensuremath{{\rm e}^{#1}}}

\newcommand{\der}{\ensuremath{{\rm d}}}

\newcommand{\erf}[1]{\ensuremath{{\rm erf}\left(#1\right)}}

\newcommand{\eqn}[1]{equation~\eqref{#1}}
\newcommand{\eqns}[1]{equations~\eqref{#1}}
\newcommand{\ph}[1]{\phantom{#1}}

\newcommand{\be}{\begin{equation}}
\newcommand{\ee}{\end{equation}}
\newcommand{\Cal}[1]{\ensuremath{\mathcal{#1}}}

\title[Analytical tidal fields]
      {Analytical halo models of cosmic tidal fields} 
\date{draft}

\author[A. Paranjape]{
Aseem Paranjape$^{1}$\thanks{E-mail: aseem@iucaa.in}\\
 $^1$ Inter-University Centre for Astronomy \& Astrophysics,
      Ganeshkhind, Post Bag 4, Pune 411007, India
}

\begin{document}

\label{firstpage}
\pagerange{\pageref{firstpage}--\pageref{lastpage}}

\maketitle 

\begin{abstract}
\noindent
The non-linear cosmic web environment of dark matter haloes plays a major role in shaping their growth and evolution, and potentially also affects the galaxies that reside in them. 
We develop an analytical (halo model) formalism to describe the tidal field of anisotropic halo-centric density distributions, as characterised by the halo-centric tidal tensor $\avg{T_{ij}}(<R)$ spherically averaged on scale $R\sim4R_{\rm vir}$ for haloes of virial radius $R_{\rm vir}$. 
 We focus on axisymmetric anisotropies, which allows us to explore simple and intuitive toy models of (sub)halo configurations that exemplify some of the most interesting anisotropies in the cosmic web. 
 We build our models around the spherical Navarro-Frenk-White (NFW) profile after describing it as a Gaussian mixture, which leads to almost fully analytical expressions for the `tidal anisotropy' scalar $\alpha(<4R_{\rm vir})$ extracted from the tidal tensor. 
 Our axisymmetric examples include (i) a spherical halo at the axis of a cylindrical filament, (ii) an off-centred satellite in a spherical host halo and (iii) an axisymmetric halo. 
 Using these, we demonstrate several interesting results. 
 For example, the tidal tensor at the axis of a pure cylindrical filament gives $\alpha^{\rm (fil)}(<R)=1/2$ exactly, for any $R$. 
 Also, $\alpha(<4R_{\rm vir,sat})$ for a satellite of radius $R_{\rm vir,sat}$ as a function of its host-centric distance is a sensitive probe of dynamical mass loss of the satellite in its host environment. 
 Finally, we discuss a number of potentially interesting extensions and applications of our formalism that can deepen our understanding of the multi-scale phenomenology of the cosmic web. 
 \end{abstract}


 \section{Introduction}
 \label{sec:intro}
 \noindent
 The cosmic tidal environment of dark matter haloes is a key arbiter of their growth and evolution, and possibly also of the physics governing the galaxies that occupy them. 
Early work in this subject \citep{zeldovich70,doroshkevich70,sz89} has established the important role of tidal fields in the formation of the cosmic web, with subsequent numerical work \citep{bm96,bm96b,bm96c,bkp96} showing that the skeleton of the cosmic web is already in place when the bulk of formation of small-scale structure occurs \citep[see also][]{vdwb08}.
Due to the non-linearity of gravitational evolution, most of the progress in our understanding of the connection between cosmic tidal fields and the physics of haloes and galaxies has been driven by numerical simulations \citep{hoffman86,vdwb93,vhvdw93,hahn+07,hahn+07b,hahn+09,codis+12,behroozi+14,chisari+15,hbv16,bprg17,phs18a,libeskind+17,kraljic+19}.  Analytical insight, however, can play an important role in clarifying the multitude of multi-scale correlations \citep[see, e.g.,][]{han+19,rphs19} that pervade the cosmic web.

The cosmic web is generally decomposed into its component structures that include filaments and sheets (or walls), with nodes occurring at the intersections of filaments and voids filling most of the intervening volume. These structures are usually quantified either using the spatial variation of the density field or using the strength of the tidal field \citep[see, e.g.,][and references therein]{libeskind+17}. In this work, we will focus on using the cosmic tidal field as a means of classifying the cosmic web (\citealp{hahn+07}; \citealp{ForeroRomero09}; see also \citealp{cvdwj13}).

There have been several (semi-)analytical studies of tidal fields starting from the seminal work of \citet{zeldovich70}, particularly on the role of initial tides in determining late-time halo distributions \citep[see, e.g.,][]{icke73,ws79,el95,bm96,vdwb96,monaco99,smt01,sams06,desjacques:2008a,rossi12,ps13,cphs17,musso+18,djs18b}.
The evolution of the cosmic web is also predicted to be intimately tied with the acquisition of angular momentum by haloes and consequently the galaxies that inhabit them. Although not the focus of this work, it is worth mentioning that tidal torque theory \citep{hoyle51,peebles69,white84} explicitly models the effect of the tidal environment on evolving proto-halo patches, thus imbibing them with angular momentum, both in the linear and non-linear regime \citep[e.g.,][]{lp00,pdh02,pdh02b,aragoncalvo07b,hahn+07,jvdw09,schaefer09}.
In parallel, there is also a well-developed analytical framework for studying spherical haloes and their spatial distribution \citep[the halo model, see][for a review]{cs02}.
However, the non-linear analytical modelling of small-scale, late-time cosmic web environments remains largely unexplored. 

To this end, in this work we develop the analytical framework needed to describe the key variable of interest, namely, the halo-centric tidal tensor $T$ (defined below) spherically averaged with a smoothing scale $R$ proportional to the halo virial radius $R_{\rm vir}$. The expressions for the smoothed tensor $T(<R)$ can then be reduced to the scalar halo-centric `tidal anisotropy' $\alpha(<4R_{\rm vir})$ (see below) which has emerged as a particularly useful indicator of the local cosmic web environment of the halo. 
 This quantity correlates tightly with the large-scale environment (or linear halo bias $b_1$) of dark matter haloes over a wide range of halo mass \citep{phs18a}. It also correlates tightly with a number of internal halo properties such as halo concentration, the anisotropy of the mass and velocity ellipsoid tensors, the anisotropy of velocity dispersion and halo spin \citep{rphs19}. In fact, \citet{rphs19} showed that the assembly (or secondary) bias, i.e., the correlation $c\leftrightarrow b_1$ at fixed halo mass \citep{st04,gsw05,jsm07,fw10} seen in \emph{all} of these internal properties $c$ can be statistically explained as arising from the two fundamental correlations $\alpha\leftrightarrow b_1$ and $\alpha\leftrightarrow c$. It is therefore of considerable interest to gain analytical insights into the behaviour of $\alpha(<4R_{\rm vir})$ in different cosmic web environments.

We will see that the tidal tensor formalism we develop considerably simplifies for an axisymmetric unsmoothed density field. We will therefore apply this restricted, axisymmetric version of our analytical framework to three toy models: (i) a halo at the axis of a filament, (ii) a satellite in a spherical halo and (iii) an axisymmetric halo, all built around the spherical NFW profile \citep*{nfw96,nfw97}. These examples are chosen to minimally illustrate a few of the most relevant anisotropies that are expected to affect (sub)halo populations in the cosmic web, and represent a trade-off between analytical simplicity and cosmological relevance. 
 Throughout, we adopt a `halo modeller' point of view, in that we aim to describe static cosmological configurations rather than predict their dynamical evolution. Nevertheless, we will see that our models can, in principle, be used to place interesting constraints on dynamical models as well.

 Despite the restriction to axisymmetric anisotropies, the integrals defining the tidal tensor do not have closed form expressions, in general. We therefore introduce a high-accuracy Gaussian mixtures description of the NFW profile. While increasing the complexity of the profile itself, this allows for expressions for the tidal tensor that are either in closed form  or involve straightforward 1-dimensional numerical integrals, in all the cases we study. We perform the Gaussian mixture fit using an implementation of a non-negative least squares algorithm in standard numerical libraries. Calculating the full tidal tensor is then essentially reduced to a trivial numerical book-keeping exercise in combining the contributions from individual Gaussian components.

 The paper is organised as follows. Section~\ref{sec:TT} sets up the tidal tensor formalism in spherical polar coordinates, focusing on spherically averaged quantities described using spherical harmonics. Section~\ref{sec:spherical} presents the Gaussian mixture description of the NFW profile. Section~\ref{sec:applicn} applies the resulting framework to the examples mentioned above and we conclude in section~\ref{sec:conclude}. The Appendices present details of calculations used in the main text.
 Unless stated otherwise, for illustrative purposes we will use $c_{\rm vir} = 7\,(R_{\rm vir}/1\Mpch)^{-0.4}$ for the halo concentration of the NFW profile, which is approximately consistent with the relation calibrated by \citet{bullock+01} using haloes in $N$-body simulations.\footnote{It is straightforward to incorporate more up-to-date calibrations such as those presented by \citet{dk15}; however, this will not alter any of our conclusions.}

 \section{Analytical formalism: tidal tensor}
 \label{sec:TT}
 \noindent
 In this section we derive our main formal results, namely, expressions for the spherically averaged halo-centric tidal tensor (defined below). Although the formalism can be developed in full generality, for analytical simplicity we eventually focus attention on the tidal tensor derived from an axisymmetric field.

 \subsection{Basic definitions}
 \label{subsec:basicTT}
 \noindent
 The primary quantity of our interest is the dimensionless tidal tensor $T(\mb{r})$ at location \mb{r} which can be written in coordinate invariant form as
 \be
 T(\mb{r}) = \left(\nabla \otimes \nabla\right)\, \psi(\mb{r})\,,
 \label{eq:TTcoordinvar}
 \ee
 where $\otimes$ denotes a direct product, $\nabla$ is the gradient operator and $\psi(\mb{r})$ is the normalised gravitational potential which obeys the Poisson equation
 \be
 \nabla^2\psi(\mb{r}) = \delta(\mb{r})\,,
 \label{eq:Poisson}
 \ee
 where 
 \be
 \delta(\mb{r}) \equiv \Delta(\mb{r})-1 = \rho(\mb{r})/\bar\rho-1
 \label{eq:deltadef}
 \ee
 is the matter density contrast (and is also consequently the trace of the tidal tensor). Typically, one considers spherically averaged versions of $\Delta$ and $\psi$. For any field $f(\mb{r})$ with Fourier transform coefficients $f_{\mb{k}}$, the smoothed field can be written as
 \be
 f(\mb{r};<R) = \Cal{F}\left[f_{\mb{k}}\,W(\mb{k};R)\right] 
 \label{eq:smoothdef}
 \ee
 where $W(\mb{k};R) = W(kR)$ is the Fourier transform of the normalised spherical smoothing window and $R$ is the smoothing radius. E.g., the Gaussian window routinely employed in simulations would have $W(kR) = \e{-k^2R^2/2}$. Below, we will exclusively consider spherical tophat averaging, for which $W(kR) = 3j_1(kR)/(kR)$ with $j_1(x)$ a spherical Bessel function. We  will drop the explicit dependence on location and/or smoothing scale in $f(\mb{r};<R)$ whenever no confusion can arise. 

 The (smoothed) tidal tensor can be diagonalised at each location; we denote its three eigenvalues as
 \be
 \lambda_1 \leq \lambda_2 \leq \lambda_3\,,
 \label{eq:eigvalorder}
 \ee
 and we have $\Delta-1 = \lambda_1+\lambda_2+\lambda_3$. The eigenvalues can be combined into the `tidal anisotropy' $\alpha$ defined as
 \be
 \alpha \equiv \sqrt{q^2} / \Delta\,,
 \label{eq:alphadef}
 \ee
 with 
 \be
 q^2 \equiv \frac12\left[(\lambda_3-\lambda_1)^2 + (\lambda_3-\lambda_2)^2 + (\lambda_2-\lambda_1)^2\right]\,.
 \label{eq:q2def}
 \ee
For the reasons discussed in the Introduction, it is of considerable interest to understand the properties of $\alpha(\mb{r}_{\rm halo};<4R_{\rm vir})$ for a halo of virial radius $R_{\rm vir}$ located at $\mb{r}_{\rm halo}$ in a variety of cosmic web environments and halo configurations. We will do so below using toy models which, despite being considerably simplified, yield valuable insight into the nature of the anisotropic cosmic tidal field.

 \subsection{Tidal tensor in polar coordinates}
 \label{subsec:TTpol}
 \noindent
 We are interested in describing spherically averaged tidal fields derived from anisotropic density distributions centred on halo locations. It is therefore convenient to perform all calculations using spherical polar coordinates $\{r,\theta,\phi\}$ along with spherical harmonic multipole expansions. We set up this formalism for the tidal tensor next. To start with, let us obtain the spherical polar components of $T$ (equation~\ref{eq:TTcoordinvar}) in spherical coordinates. We will then use these to describe the spherically averaged \emph{Cartesian} components of $T$, which will lead in a straightforward manner to expressions for $\alpha(<R)$.

 Denoting the spherical polar orthonormal basis vectors as $\{\mb{e}_\alpha\}= \{\mb{e}_r,\mb{e}_\theta,\mb{e}_\phi\} = \{\hat r,\hat\theta,\hat\phi\}$ and partial derivatives as $\p/\p r = \p_r$, etc., and writing the tidal tensor as $T = T_{\alpha\beta}\left(\mb{e}_\alpha\otimes\mb{e}_\beta\right)$, the spherical polar components $T_{\alpha\beta}$ can then be expressed in terms of spherical polar derivatives of $\psi$ using the expression
 \be
 \nabla = \hat r\,\p_r + \frac1r\,\hat\theta\,\p_\theta + \frac1{r\,s_\theta}\,\hat\phi\,\p_\phi \equiv \hat r\,\p_r + \frac1r\,\nabla_\Omega\,,
 \label{eq:gradpol}
 \ee
 along with derivative identities relating the basis vectors. Here and below, we abbreviate $\sin(\theta)=s_\theta$ and $\cos(\theta)=c_\theta$ for any angle $\theta$. Appendix~\ref{app:TTpol} shows that this results in the expressions
 \begin{align}
 T_{rr} &= \p_r^2\psi \quad;\quad T_{\theta\theta} = \frac1r\,\p_r\psi + \frac1{r^2}\,\p_\theta^2\psi\notag\\
 T_{\phi\phi} &= \frac1r\,\p_r\psi + \frac1{r^2}\,\frac{c_\theta}{s_\theta}\,\p_\theta\psi + \frac1{(r\,s_\theta)^2}\,\p_\phi^2\psi\notag\\
 T_{r\theta} &= \frac1{r^2}\left(r\p_r-1\right)\,\p_\theta\psi \quad;\quad T_{r\phi} = \frac1{r^2\,s_\theta}\left(r\p_r-1\right)\,\p_\phi\psi \notag\\
 T_{\theta\phi} &= \frac1{r^2\,s_\theta}\left(\p_\theta - \frac{c_\theta}{s_\theta}\right)\,\p_\phi\psi\,.
 \label{eq:TTpol}
 \end{align}
 Appendix~\ref{app:TTpol} also relates these spherical polar components $T_{\alpha\beta}(\mb{r})$ to the more familiar Cartesian components $T_{ij}(\mb{r})$, $i,j\in\{x,y,z\}$ using a local rotation of basis vectors at \mb{r}. These are summarised in \eqn{eq:Tpol-to-Txyz}.

 \subsection{Spherical averaging}
 \label{subsec:sphavg}
 \noindent
 We are interested in spherically averaging the halo-centric tidal tensor \eqn{eq:TTpol}. In general, for any quantity $T(\mb{r})$ expressed in spherical polar coordinates, smoothing with a spherical tophat of radius $R$ can be written as an angular average followed by a radial average, i.e.,
 \be
 \avg{T}(<R)\equiv \avg{\avg{T}_\Omega}_R
 \label{eq:sphavg-sphpol}
 \ee
 where the angular average is defined by
 \begin{align}
 \avg{f}_\Omega &\equiv \int\frac{\der\Omega}{4\pi}\,f(\mu,\phi) \notag\\
 &= \int_{-1}^1\frac{\der\mu}{2}\int_0^{2\pi}\frac{\der\phi}{2\pi}\,f(\mu,\phi)\notag\\
 &\equiv \avg{\avg{f}_\phi}_\mu\,,
 \label{eq:angavg}
 \end{align}
 where we defined
 \be
 \mu\equiv \cos(\theta)\,,
 \label{eq:mudef}
 \ee
 and the average over radial shells is
 \be
 \avg{g}_R \equiv \frac{3}{R^3}\int_0^R\der r\,r^2\,g(r)\,.
 \label{eq:radavg}
 \ee
 In the following, we will also use the notation $g(<R)$ to denote the average \eqref{eq:radavg} of any radial function $g(r)$.

 \subsection{Multipole expansions}
 \label{subsec:multipole}
 \noindent
 Writing $\mb{r} = r\,\hat r(\mu,\phi)$, we can expand the \emph{unsmoothed}, halo-centric gravitational potential $\psi(\mb{r})$ and local overdensity $\Delta(\mb{r})$ in spherical harmonics as
 \begin{align}
 \psi(\mb{r}) &= \sum_{\ell=0}^{\infty}\sum_{m=-\ell}^{\ell} \, \psi_{\ell m}(r)\,Y^m_\ell(\hat r)
 \label{eq:psi-multipoleexpn}\\
 \Delta(\mb{r})-1 &= \sum_{\ell=0}^{\infty}\sum_{m=-\ell}^{\ell} \, \Delta_{\ell m}(r)\,Y^m_\ell(\hat r)\,,
 \label{eq:Delta-multipoleexpn}
 \end{align}
 (note the $-1$ on the l.h.s. of equation~\ref{eq:Delta-multipoleexpn}) where $Y^m_\ell(\hat r)$ are spherical harmonics. Appendix~\ref{app:multipole} recapitulates some useful properties of the $Y^m_\ell(\hat r)$ and Legendre polynomials $P_\ell(\mu)$. 

 Using the multipole expansion of the Green's function of the Laplacian $\nabla^2$, the Poisson equation can be integrated in spherical polar coordinates to obtain the solution \citep[][section 2.4]{binney-tremaine-GalDyn}
 \be
 \psi_{\ell m}(r) = -\frac{1}{2\ell+1}\left[\frac1{r^{\ell+1}}\Cal{I}_{\ell m}(r) + r^\ell\,\Cal{O}_{\ell m}(r)\right]\,,
 \label{eq:Poissonsoln-1}
 \ee
 where we defined
 \begin{align}
 \Cal{I}_{\ell m}(r) &= \int_0^r\der u\,u^2\,u^\ell\,\Delta_{\ell m}(u)\notag\\
 \Cal{O}_{\ell m}(r) &= \int_r^\infty\der u\,u^2\,\frac1{u^{\ell+1}}\,\Delta_{\ell m}(u)\,.
 \label{eq:Poissonsoln-2}
 \end{align}
 Armed with a model for the anisotropy of the density distribution given by the functions $\Delta_{\ell m}(r)$, we can then use equations~\eqref{eq:psi-multipoleexpn} and~\eqref{eq:Poissonsoln-1} to evaluate the tidal tensor components in \eqn{eq:TTpol} at any location. 

 \subsection{Axisymmetric model}
 \label{subsec:axisymm}
 \noindent
 In principle, one could now sequentially perform the azimuthal, polar and radial averages of the tidal tensor and then calculate $\alpha$ using \eqns{eq:alphadef}, \eqref{eq:q2def} and \eqref{eq:I1I2}.  Appendix~\ref{app:azavg} presents the azimuthal average of the tidal tensor; this shows that \emph{all} Cartesian components of the tidal tensor are, in general, non-vanishing, which makes the subsequent steps rather involved. To simplify the discussion, in the following we will therefore \emph{restrict attention to axisymmetric potentials described by $m=0$ but generic $\ell$, for which the angle-averaged tidal tensor is diagonal in the Cartesian frame} (equation~\ref{eq:<Tij>_Omega:ell0} in Appendix~\ref{app:axisymm}), and will return to a fuller discussion of generic potentials in future work.

 Since the tidal tensor $T$ is linear in the potential $\psi$, which is itself linear in overdensity $\Delta$, we can analyse each multipole moment separately to begin with. For a pure multipole $\ell$, restricting to $m=0$, we have 
 \be
 \psi(\mb{r}) = N_{\ell0}\,P_\ell(\mu)\,\psi_{\ell0}(r)\,,
 \label{eq:psiell0}
 \ee
 where the normalisation constant $N_{\ell m}$ is defined in \eqn{eq:Ylm-norm} and included here for convenience.

 Appendix~\ref{app:axisymm} also shows that the angle-averaging the Cartesian components of the tidal tensor derived from \eqn{eq:psiell0} kills all multipoles except $\ell=0$ and $\ell=2$, leading to
 \begin{align}
 \avg{T_{xx}}_\Omega &= \frac13\left[\avg{\Delta}_\Omega(r)-1\right]\delta_{\ell,0} + \frac12\,t(r)\,\delta_{\ell,2} = \avg{T_{yy}}_\Omega\notag\\
 \avg{T_{zz}}_\Omega &= \frac13\left[\avg{\Delta}_\Omega(r)-1\right]\delta_{\ell,0} - t(r)\,\delta_{\ell,2}\,.
 \label{eq:<Tij>_Omega:ell0-explicit}
 \end{align}
 where 
 \be
 \avg{\Delta}_\Omega(r)-1=N_{00}\Delta_{00}=N_{00}r^{-2}\p_r(r^2\p_r\psi_{00})\,,
 \label{eq:<Delta>_Omega}
 \ee
 and we defined
 \be
 t(r) \equiv -\frac13\sqrt{\frac{1}{5\pi}}\left[\psi_{20}^{\prime\prime}+\frac5r\,\psi_{20}^\prime+\frac3{r^2}\,\psi_{20}\right]\,.
 \label{eq:tdef}
 \ee
 Performing the radial average and generalising to an arbitrary axisymmetric density field satisfying 
 \begin{align}
 \Delta(\mb{r}) - 1 &=\sum_{\ell=0}^\infty\,N_{\ell 0}P_\ell(\mu)\Delta_{\ell0}(r)\,, 
 \label{eq:Delta-genericaxisymm}\\
 \Delta_{\ell0}(r) &= 4\pi\,N_{\ell0}\int_{-1}^1\frac{\der\mu}{2}\,P_\ell(\mu)\left(\Delta(\mb{r})-1\right)\,,
 \label{eq:Delta-ell0}
 \end{align}
 the Cartesian components of the spherically averaged tidal tensor can be written as
 \be
 \avg{T_{ij}}(<R) = {\rm diag}\left\{\lambda_+,\lambda_+,\lambda_-\right\}
 \label{eq:<Tij>(<R):axisymm}
 \ee
 with
 \begin{align}
 \lambda_+ &\equiv \frac13\left[\avg{\Delta}(<R)-1\right] + \frac12 t(<R)\,, \notag\\
 \lambda_- &\equiv \frac13\left[\avg{\Delta}(<R)-1\right] - t(<R)\,,
 \label{eq:lambda+-def}
 \end{align}
 where $t(<R)$ is the radial average of $t(r)$ (equation~\ref{eq:radavg}) and can be written using straightforward integration by parts as
 \begin{align}
 t(<R) &= -\frac1{\sqrt{5\pi}}\left(\frac1{r^4}\,\p_r\left(r^3\psi_{20}(r)\right)\right)\bigg|_{r=R}\notag\\
 &= \frac1{\sqrt{5\pi}}\,\int_R^\infty\frac{\der r}{r}\,\Delta_{20}(r)\,,
 \label{eq:t(<R)}
 \end{align}
 with the second equality following from \eqns{eq:Poissonsoln-1} and~\eqref{eq:Poissonsoln-2}, and we have
 \begin{align}
 \avg{\Delta}(<R)-1 &= \frac{1}{\sqrt{4\pi}}\,\Delta_{00}(<R)\notag\\
 &=\frac{3}{\sqrt{4\pi}}\left(\frac1r\,\p_r\psi_{00}(r)\right)\bigg|_{r=R}\,.
 \label{eq:delta(<R)}
 \end{align}
 Using \eqn{eq:q2def}, we then have
 \be
 q^2(<R) = \left(\lambda_+-\lambda_-\right)^2 = \left(\frac{3}{2}t(<R)\right)^2\,,
 \label{eq:q2(<R):axisymm}
 \ee
 and the tidal anisotropy $\alpha(<R)$ in \eqn{eq:alphadef} for a generic axisymmetric density field is thus given by
 \begin{align}
 \alpha(<R) &= \frac32\,\frac{\left|t(<R)\right|}{\avg{\Delta}(<R)}\,.
 \label{eq:alpha(<R):summary}
 \end{align}
\emph{Equations~\eqref{eq:t(<R)}, \eqref{eq:delta(<R)} and~\eqref{eq:alpha(<R):summary}, valid for any axisymmetric density distribution \eqref{eq:Delta-genericaxisymm}, form the core result of this work.}

 \subsection{Special cases}
 \label{subsec:asides}
 \noindent
 Before exploring applications of these results, we pause to consider two interesting off-shoots of our calculations above.

 \subsubsection{Density Hessian}
 \label{subsubsec:hessian}
 \noindent
 It is interesting to compare the calculation of the tidal tensor above with the closely related density Hessian 
 \be
 H \equiv (\nabla\otimes\nabla)(\Delta-1)\,.
 \label{eq:hessiandef}
 \ee
The spherically averaged density Hessian is also diagonal in the Cartesian frame in the axisymmetric model. We can directly apply the formalism above to write
 \be
 \avg{H_{ij}}(<R) = {\rm diag}\left\{\Lambda_+,\Lambda_+,\Lambda_-\right\}\,,
 \label{eq:<Hij>(<R):axisymm}
 \ee
 with
 \begin{align}
 \Lambda_+ &\equiv \frac13\chi(<R) + \frac12 \tau(<R)\,, \notag\\
 \Lambda_- &\equiv \frac13\chi(<R) - \tau(<R)\,,
 \label{eq:Lambda+-def}
 \end{align}
 where
 \begin{align}
 \chi(<R)&=\frac{3}{\sqrt{4\pi}}\left(\frac1r\,\p_r\Delta_{00}(r)\right)\bigg|_{r=R}\,,
 \label{eq:gammadef}\\
 \tau(<R) &= -\frac1{\sqrt{5\pi}}\left(\frac1{r^4}\,\p_r\left(r^3\Delta_{20}(r)\right)\right)\bigg|_{r=R}\,.
 \label{eq:taudef}
 \end{align}
 In other words, the spherically averaged tidal tensor and density Hessian are perfectly aligned in the axisymmetric model. Since this was a direct consequence of assuming that only $m=0$ contributes (see, e.g., equation~\ref{eq:Txyz-phiavg}), \emph{any misalignment of the tidal tensor and density Hessian must be directly connected to departures from axisymmetry of the density distribution.} In other words, the misalignment angle between these two \emph{smoothed} tensors can be thought of as a direct measurement of triaxiality of the \emph{unsmoothed} field.

 \subsubsection{Perfect spherical symmetry}
 \label{subsubsec:sphsymm}
 \noindent
 It is also instructive to recover the tidal tensor for a perfectly spherically symmetric density field from the formalism above. When $\Delta(\mb{r}) = \Delta(r)$, only the monopole $\ell=0$, $m=0$ survives in \eqn{eq:Poissonsoln-1} so that $\psi(\mb{r}) = \psi(r)$ satisfies $r^{-2}\p_r(r^2\p_r\psi(r))=\Delta(r)-1$, and \eqn{eq:TTcoordinvar} consequently becomes the diagonal form
 \begin{align}
 T &= \left(\hat r\otimes\hat r\right)\,\p_r^2\psi + \left(\hat \theta\otimes\hat \theta + \hat \phi\otimes\hat \phi\right)\,\frac1r\,\p_r\psi \notag\\
 &= \frac13\,\mb{1}\left( \frac3r\,\p_r\psi\right) + \left(\hat r\otimes\hat r\right)\left(\p_r^2\psi - \frac1r\,\p_r\psi\right)\notag\\
 &= \frac13\,\mb{1}\,\left(\Delta(r)-1\right) + \left(\frac13\,\mb{1} - \hat r\otimes\hat r\right)\left(\Delta(<r) - \Delta(r)\right)\,,
 \label{eq:TTcoordinvar-sphsym}
 \end{align}
 where $\mb{1}=\hat r\otimes\hat r + \hat \theta\otimes\hat \theta + \hat \phi\otimes\hat \phi$ is the identity matrix. This agrees with \citet{ps13}, with the first term being the isotropic part proportional to the differential density contrast $\Delta(r)-1$ and the second term being the trace-free contribution.

 \begin{figure*}
 \centering
 \includegraphics[width=0.45\textwidth]{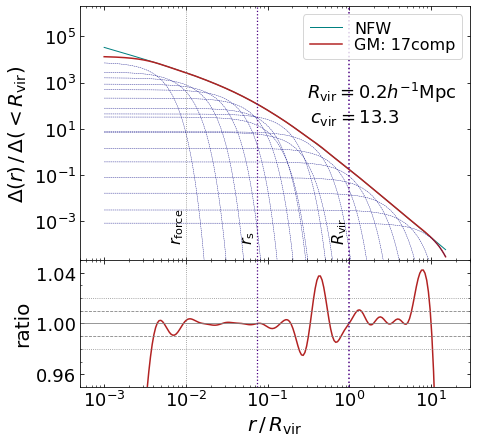}
 \includegraphics[width=0.45\textwidth]{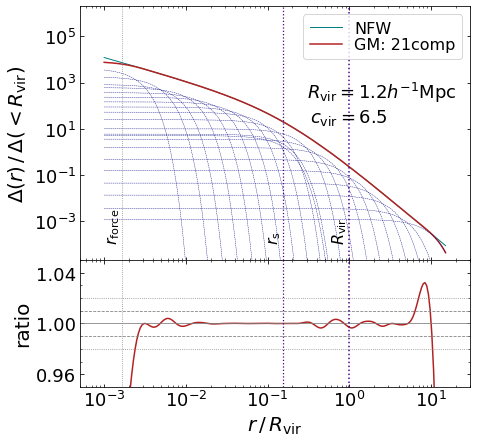}
 \caption{Gaussian mixture approximations for the NFW profile for two combinations of virial radius $R_{\rm vir}$ and concentration $c_{\rm vir}$, corresponding to a small ($m_{\rm vir}\simeq2\times10^{12}\Mh$; \emph{left panels}) and a large halo ($m_{\rm vir}\simeq4\times10^{14}\Mh$; \emph{right panels}). 
 \emph{(Top row:)} Cyan curves in each panel show the original NFW profile \eqref{eq:NFWprofile}, normalised to enclose unit density inside $R_{\rm vir}$ and shown as a function of halo-centric distance $r$ in units of $R_{\rm vir}$.
 Red curves show the corresponding best fitting Gaussian mixture \eqref{eq:NFWgaussmix} with the number of components indicated in the respective legend.
 Thin purple curves show individual Gaussian components.
 \emph{(Bottom row:)} Ratio of the Gaussian mixture to the original NFW profile.
 The fits were performed using a constrained non-linear least squares method as described in section~\ref{subsec:gm-nfw} and Appendix~\ref{app:GMdetails}, setting the ``force resolution'' $r_{\rm force}=2\kpch$. 
 Vertical dotted lines in each panel indicate the values (from left to right) of $r_{\rm force}$, scale radius $r_{\rm s}=R_{\rm vir}/c_{\rm vir}$ and $R_{\rm vir}$.
 For these examples, we see sub-percent deviations between the fit and the original profile over nearly the entire dynamic range explored, except for local peaks and troughs reaching inaccuracies of a few per cent. See Figure~\ref{fig:error-gmNFW} for a comprehensive summary of errors in the fits over the parameter space of $\{R_{\rm vir},c_{\rm vir}\}$.}
 \label{fig:gaussmixnfw}
 \end{figure*}

 It is also interesting to note that the angle average of this tensor $\avg{T}_\Omega$ completely kills the trace-free part. This is easily seen by converting \eqn{eq:TTcoordinvar-sphsym} to Cartesian components using \eqn{eq:Tpol-to-Txyz} and then using the identities \eqref{eq:phiavgs} and $\avg{\mu^2}_\mu=1/3$, which leads to $\avg{T}_\Omega=\mb{1}\,(\Delta(r)-1)/3$, which agrees with \eqn{eq:<Tij>_Omega:ell0-explicit} for the case where only the monopole $\ell=0$ contributes. Radially averaging this gives $\avg{T}(<R)=\mb{1}\,(\Delta(<R)-1)/3$; thus, the spherical average of a spherically symmetric tidal tensor is isotropic. \emph{Any anisotropy in the spherically averaged halo-centric tidal tensor is therefore a direct consequence of anisotropy in the unsmoothed density field.}

 \section{Analytical formalism: spherical profiles}
 \label{sec:spherical}
 \noindent
 In this section, we develop a Gaussian mixtures approach to describing spherically symmetric density profiles, which considerably simplifies the application of the tidal tensor formalism to interesting halo model configurations. We remind the reader that we are interested in developing a descriptive framework rather than tracking dynamical evolution, and therefore assume pre-existing, non-linearly evolved density profiles for the halo interior, wherever needed.

 \subsection{The NFW profile}
 \label{subsec:nfw}
 \noindent
 As our base spherical profile, we will use the 2-parameter isotropic NFW profile \citep{nfw96,nfw97}, which provides a simple and accurate description of the spherically averaged dark matter density profile of haloes identified in $N$-body simulations. The profile is given by
 \be
 \Delta_{\rm NFW}(r|R_{\rm vir},c_{\rm vir}) =  \frac{\Delta_{\rm vir}}{3}\,\frac{c_{\rm vir}^3}{f(c_{\rm vir})}\,\frac1{(r/r_{\rm s})\left(1+r/r_{\rm s}\right)^2}\,,
 \label{eq:NFWprofile}
 \ee
 where $R_{\rm vir}$ and $c_{\rm vir}$ are, respectively, the virial radius and concentration of the halo, in terms of which the scale radius $r_{\rm s}$ satisfies 
 \be
 r_{\rm s} = R_{\rm vir}/c_{\rm vir}\,,
 \label{eq:rsrvircvir}
 \ee
 and we defined the function 
 \be
 f(c) \equiv \int_0^c\frac{\der y\,y}{(1+y)^2} = \ln(1+c) - \frac{c}{1+c}\,.
 \label{eq:fdef}
 \ee
 The profile \eqref{eq:NFWprofile} is normalised so as to enclose a mass $m_{\rm vir}=(4\pi/3)R_{\rm vir}^3\Delta_{\rm vir}\bar\rho$ inside the virial radius, i.e., 
 \be
 \Delta_{\rm NFW}(<R_{\rm vir}|R_{\rm vir},c_{\rm vir}) = \Delta_{\rm vir}\,.
 \label{eq:NFWnorm}
 \ee

 \subsection{Gaussian mixtures description}
 \label{subsec:gm-nfw}
 \noindent
 We wish to explore anisotropic (axisymmetric) models built around the spherical NFW profile~\eqref{eq:NFWprofile}. Despite its simplicity and the fact that analytical results exist in closed form for many associated properties such as the gravitational potential, velocity dispersion, etc. \citep[see, e.g.,][]{shds01}, the integrals involved in computing the spherical harmonic coefficients $\Delta_{\ell0}(r)$ for the axisymmetric model (equation~\ref{eq:Delta-ell0})
 do not typically have closed form expressions. However, a spherically symmetric \emph{Gaussian} profile $\propto\e{-r^2/R_\ast^2}$ for some scale radius $R_\ast$, does in fact lead to almost fully analytical expressions for $\avg{\Delta}(<R)$ and $t(<R)$ for the models we explore.

 \begin{figure*}
 \centering
 \includegraphics[width=0.45\textwidth]{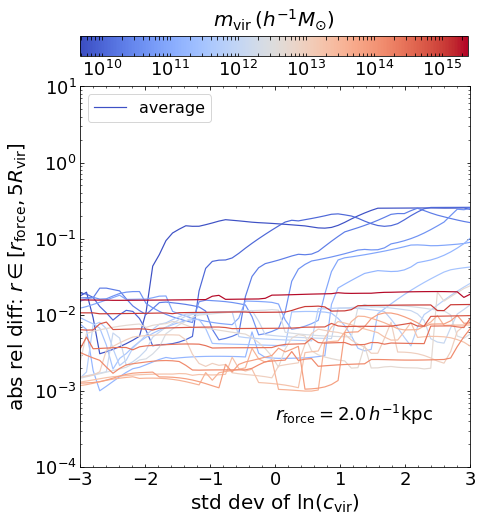}
 \includegraphics[width=0.45\textwidth]{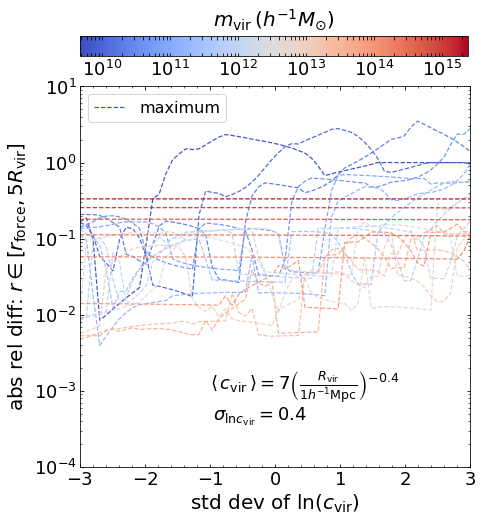}
 \caption{Average \emph{(left panel)} and maximum \emph{(right panel)} of the absolute relative difference $\epsilon\equiv|\Delta_{\rm GM}(r)/\Delta_{\rm NFW}(r)-1|$ between the NFW profile \eqref{eq:NFWprofile} and its Gaussian mixture approximation \eqref{eq:NFWgaussmix} measured in the range $r\in[r_{\rm force},5R_{\rm vir}]$ for a range of values of halo virial mass $m_{\rm vir}$ (indicated by the colour bar) and as a function of the standardised scatter of halo concentration $c_{\rm vir}$ away from the median value at fixed halo mass, assuming a lognormal distribution with width $\sigma_{\ln c_{\rm vir}}=0.4$ \citep{wechsler+02}. For the median concentration we use $\avg{c_{\rm vir}}=7\,(R_{\rm vir}/1\Mpch)^{-0.4}$ which is approximately consistent with the relation calibrated by \citet{bullock+01} (using more up-to-date calibrations does not change the results). The average value of $\epsilon$ remains below $\sim1\%$ for all but the smallest and largest haloes we consider, at nearly all concentrations. The maximum is similarly better than $\sim20\%$ except for the smallest and largest haloes. Note that the range $[r_{\rm force},5R_{\rm vir}]$ spans nearly 3 orders of magnitude in halo-centric distance for $m_{\rm vir}\gtrsim10^{14}\Mh$. See also Figure~\ref{fig:gaussmixnfw}.}
 \label{fig:error-gmNFW}
 \end{figure*}

 In the following, therefore, we approximate the original isotropic NFW function with a Gaussian mixture,
 \be
 \Delta_{\rm NFW}(r|R_{\rm vir},c_{\rm vir}) = \sum_j\,w_j\,\Delta_{\rm G}(r|\sigma_j,r_{\rm s})\,,
 \label{eq:NFWgaussmix}
 \ee
 with $r_{\rm s}$ given by \eqn{eq:rsrvircvir} and where 
 \be
 \Delta_{\rm G}(r|\sigma_j,r_{\rm s}) = \Delta_j\,\e{-r^2/(2r_{\rm s}^2\sigma_j^2)}\,,
 \label{eq:singlegauss}
 \ee
 with pre-decided (dimensionless) widths $\{\sigma_j\}$ and normalisations $\{\Delta_j\}$. Our choice of widths $\sigma_j$ and the number of components is described in Appendix~\ref{app:GMdetails}. The weights $\{w_j\}$ are determined by a non-negative least squares calculation \citep{lh95} subject to the constraint
 \be
 \sum_j\,w_j = 1\,.
 \label{eq:wkconstraint}
 \ee
Appendix~\ref{app:GMdetails} summarises our implementation. We normalise each Gaussian component \eqref{eq:singlegauss} so as to enclose the \emph{same} mass $m_{\rm vir}$ inside $R_{\rm vir}$ as the full NFW profile, i.e., we demand $\Delta_{\rm G}(<R_{\rm vir}|\sigma_j,r_{\rm s}) = \Delta_{\rm vir}$ for each $j$, obtaining
 \be
 \Delta_j = \frac{\Delta_{\rm vir}}{3}\,\frac{c_{\rm vir}^3}{g(c_{\rm vir}|\sigma_j)}\,,
 \label{eq:gaussnorm}
 \ee
 where we defined the function
 \begin{align}
 g(c|\sigma) &\equiv \int_0^c\der y\,y^2\,\e{-y^2/(2\sigma^2)}\notag\\
 &=\sqrt{\frac\pi2}\,\sigma\,\erf{\frac{c}{\sqrt{2}\sigma}} - \sigma^2c\,\e{-c^2/(2\sigma^2)}\,.
 \label{eq:gdef}
 \end{align}
 With this choice of normalisation, the weights $\{w_j\}$ correspond to the mass fraction contributed by the respective Gaussian components.

 Figure~\ref{fig:gaussmixnfw} shows the results of the Gaussian mixture fit for two combinations of $R_{\rm vir}$ and $c_{\rm vir}$. The Gaussian mixture is naturally adapted to describing the profiles measured in actual $N$-body simulations which have finite force resolution $r_{\rm force}$ which leads to a flattening (or core) in the measured profile at $r\lesssim r_{\rm force}$. This can be trivially mimicked by the Gaussian mixture by simply dropping the appropriate number of components with the smallest $\sigma_j$ values. The fits in Figure~\ref{fig:gaussmixnfw} were constructed to safely describe the NFW profile in simulations with $r_{\rm force}\geq2\kpch$ at $z=0$. 

Figure~\ref{fig:error-gmNFW} shows the average \emph{(left panel)} and maximum \emph{(right panel)} of the absolute relative difference between the actual NFW profile and its Gaussian mixture, measured in the range $[r_{\rm force},5R_{\rm vir}]$ with $r_{\rm force}=2\kpch$, for a wide dynamic range of halo mass and concentration. We clearly obtain very high accuracy (error $\lesssim1\%$ on average) at all but the lowest and highest masses and concentrations. Note that the range $[r_{\rm force},5R_{\rm vir}]$ spans nearly 3 orders of magnitude in halo-centric distance for $m_{\rm vir}\gtrsim10^{14}\Mh$.

 We also note that the reasoning in this section applies to any spherical profile \citep[e.g.,][]{einasto65,burkert95,navarro+04}, particularly profiles with cores (see above), so that Gaussian mixtures can be used much more generally than for our specific choice of the NFW profile.

 \section{Applications}
 \label{sec:applicn}
 \noindent
 In this section, we apply the previous formalism to build anisotropic halo models for the density and tidal environment around halo locations. We study three models, one for the tidal field experienced by a halo at the centre of a cylindrical filament, the second for an off-centred satellite in a spherical host halo and the last describing the halo-centric tidal field of an axisymmetric halo. As we saw in section~\ref{subsec:axisymm}, the tidal tensor formalism simplifies \emph{considerably} for an axisymmetric unsmoothed density field. The examples we have chosen, which all involve a single special direction and are hence axisymmetric, thus represent a trade-off between analytical simplicity and cosmological relevance. Each of the examples is a toy version of more realistic configurations known or expected to exist in the cosmic web. We will also discuss potential extensions to more realistic anisotropies wherever possible.

 \subsection{Spherical halo in a filament}
 \label{subsec:filhalo}
\begin{figure}
 \centering
 \includegraphics[width=0.5\textwidth]{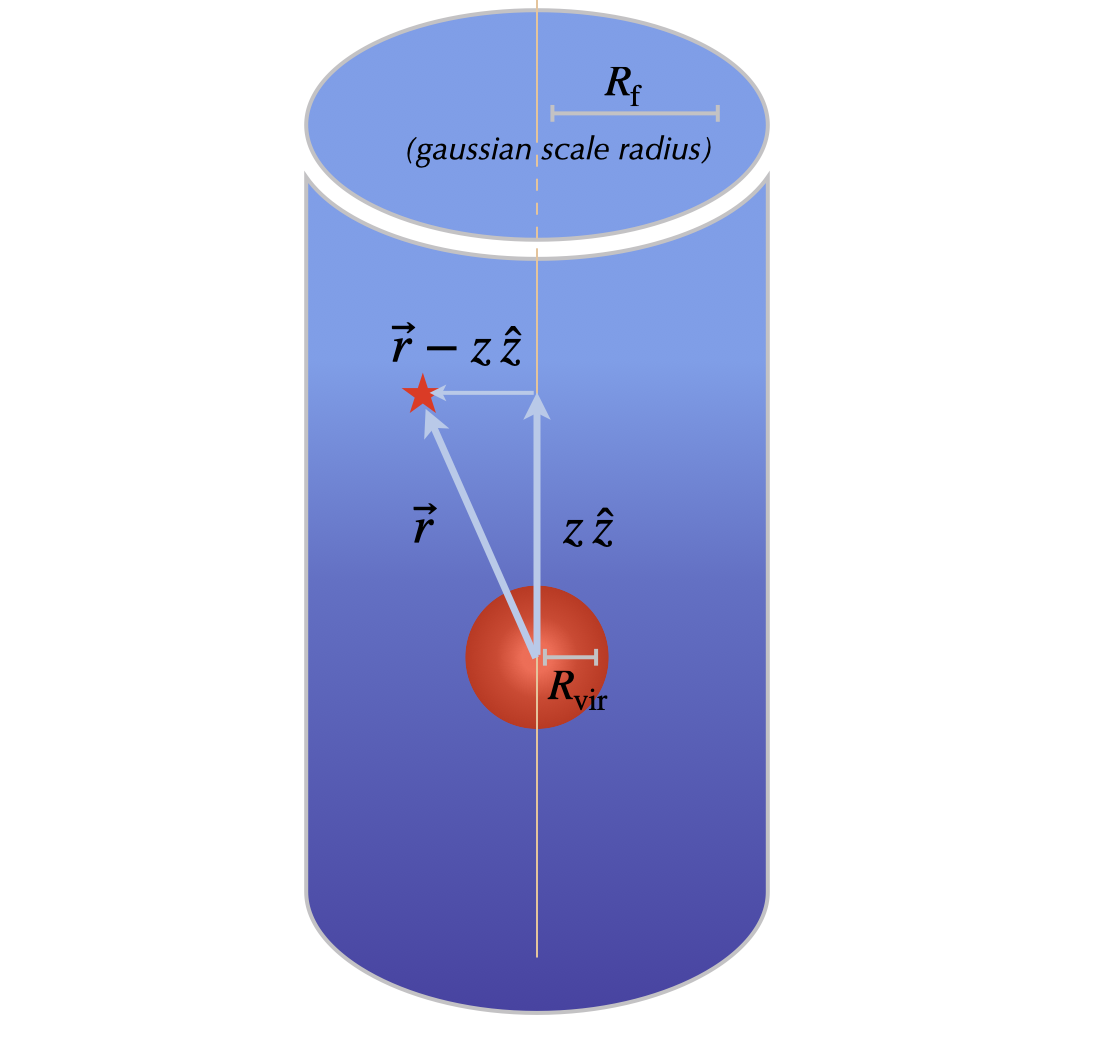}
 \caption{Illustration of a spherical NFW halo of radius $R_{\rm vir}$ located at the axis of a filament with a Gaussian profile \eqref{eq:filex-Delta^fil} with scale radius $R_{\rm f}$. The halo sees an axisymmetric tidal field due to the cylindrical density profile of the filament.}
 \label{fig:cartoon-filament}
 \end{figure}

 \noindent
 As our first example, we consider the simple case of a spherical NFW halo of radius $R_{\rm vir}$ and concentration $c_{\rm vir}$ placed \emph{exactly on the axis} of a cylindrical filament (which we align with the Cartesian $z$-axis). This configuration leads to the minimum tidal anisotropy a halo can experience in a filament, since any off-axis displacements would only serve to increase the anisotropy. Figure~\ref{fig:cartoon-filament} illustrates the situation.

 The halo-centric dark matter overdensity at location \mb{r} can be written as
 \be
 \Delta(\mb{r}) = \Delta^{\rm (fil)}(\mb{r}) + \Delta^{\rm (halo)}(\mb{r})\,,
 \label{eq:filex-Delta}
 \ee
 with the superscripts on each term on the right indicating the two contributions.

 The halo contribution $\Delta^{\rm (halo)}(\mb{r})$ is given by the spherical Gaussian mixture \eqref{eq:NFWgaussmix}. We model the filament contribution as a single Gaussian function of the perpendicular separation $r_\perp = |\mb{r}-z\hat z|=r\sqrt{1-\mu^2}$:
 \begin{align}
 \Delta^{\rm (fil)}(\mb{r}) 
 &= \frac{f_{\rm fil}\Delta_{\rm vir}}{2(1-\e{-1/2})}\,\e{-r_\perp^2/(2R_{\rm f}^2)} \notag\\
 &\equiv \Delta_{\rm f}\,\e{-r^2(1-\mu^2)/(2R_{\rm f}^2)} \,.
 \label{eq:filex-Delta^fil}
 \end{align}
 The normalisation $\Delta_{\rm f}$ is chosen such that the overdensity enclosed in a cylinder of radius $r_\perp=R_{\rm f}$ is $f_{\rm fil}\Delta_{\rm vir}$. Although not necessarily a realistic description of filamentary profiles, it should be clear that this model is trivially generalisable to arbitrary functional forms that can be described by a Gaussian mixture, by adding components with different choices of $R_{\rm f}$ and $\Delta_{\rm f}$. 

 \begin{figure*}
 \centering
 \includegraphics[width=0.45\textwidth]{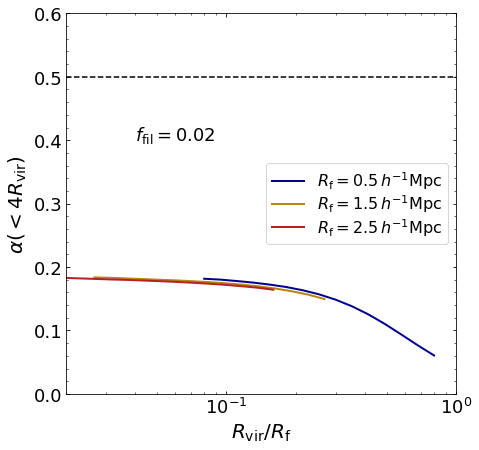}
 \includegraphics[width=0.45\textwidth]{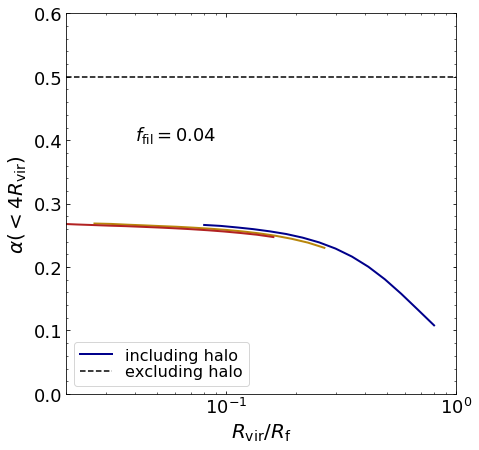}
 \caption{Tidal anisotropy $\alpha(<4R_{\rm vir})$ (equation~\ref{eq:alpha(<R):summary}) for a spherical NFW halo of virial radius $R_{\rm vir}$ placed at the axis of a cylindrical filament. The filament is chosen to have a cylindrical Gaussian density profile \eqref{eq:filex-Delta^fil} with scale radius $R_{\rm f}$ as indicated by different colours and $f_{\rm fil}=0.02\,(0.04)$ in the \emph{left (right) panel}. See Figure~\ref{fig:cartoon-filament} for an illustration. Results are shown as a function of $R_{\rm vir}/R_{\rm f}$. Solid curves show results when including the contribution of the halo profile; these are approximately universal in $R_{\rm vir}/R_{\rm f}$ as discussed in the text. Dashed horizontal line in each panel shows the value $1/2$ obtained by ignoring the halo's density and using only \eqns{eq:filex-Delta^fil(<R)} and~\eqref{eq:filex-t^fil(<R)} to calculate $\alpha$.}
 \label{fig:filhalo}
 \end{figure*}

 The corresponding multipole moments (equation~\ref{eq:Delta-ell0}) satisfy
 \begin{align}
 &\frac1{\sqrt{4\pi}}\Delta_{\ell0}^{\rm (fil)}(r) + \delta_{\ell,0} \notag\\
 &\ph{\Delta}
 = \sqrt{2\ell+1}\,\Delta_{\rm f}\int_{-1}^1\frac{\der\mu}{2}\,P_\ell(\mu)\,\e{-r^2(1-\mu^2)/(2R_{\rm f}^2)}\,.
 \label{eq:filex-Delta_ell0^fil}
 \end{align}
 These can be simplified using the relation 
 \be
 \int_{-1}^1\frac{\der\mu}{2}\,\mu^2\,\e{\mu^2a^2} = \frac1{2a}\frac{\der}{\der a}\int_{-1}^1\frac{\der\mu}{2}\,\e{\mu^2a^2}\,,
 \label{eq:muinteg-convenience}
 \ee
 and the identity
 \be
 \int_{-1}^1\frac{\der\mu}{2}\,\e{\mu^2a^2} = \int_0^1\der\mu\,\e{\mu^2a^2} = \e{a^2}\left(\frac{\Cal{D}(a)}{a}\right)\,,
 \label{eq:muinteg}
 \ee
 where $\Cal{D}(z)$ is Dawson's integral \citep[][chapter 7]{abramowitz-stegun}
 \be
 \Cal{D}(z) \equiv \e{-z^2}\int_0^z\der t\,\e{t^2}\,,
 \label{eq:Dawsondef}
 \ee
which can be evaluated using standard libraries (e.g., as {\tt scipy.special.dawsn} in SciPy).
 For convenience below, we set
 \be
 a \equiv \frac{r}{\sqrt{2}\,R_{\rm f}}\,;\quad A \equiv \frac{R}{\sqrt{2}\,R_{\rm f}}\,.
 \label{eq:aAdef}
 \ee
 A straightforward calculation then gives
 \be
 \frac1{\sqrt{4\pi}}\Delta_{00}^{\rm (fil)}(r) = \Delta_{\rm f}\,\frac{\Cal{D}(a)}{a}\,,
 \label{eq:filex-Delta_00^fil}
 \ee
 and
 \be
 \frac1{\sqrt{5\pi}}\Delta_{20}^{\rm (fil)}(r) = \frac{3\Delta_{\rm f}}{2a}\left[\frac1a-\left(\frac23+\frac1{a^2}\right)\Cal{D}(a)\right]\,.
 \label{eq:filex-Delta_20^fil}
 \ee
 Using these, the expression for $\langle{\Delta^{\rm (fil)}}\rangle(<R)$ becomes
 \be
 \langle{\Delta^{\rm (fil)}}\rangle(<R) = \frac{3\Delta_{\rm f}}{2A^2}\left(1-\frac{\Cal{D}(A)}{A}\right)\,.
 \label{eq:filex-Delta^fil(<R)}
 \ee
 Remarkably, $t^{\rm (fil)}(<R)$ can also be brought to closed form in terms of Dawson's integral; a lengthy but straightforward calculation shows that
 \begin{align}
 t^{\rm (fil)}(<R) &= \frac{3\Delta_{\rm f}}{2}\int_A^\infty\frac{\der a}{a^2} \left[\frac1a-\left(\frac23+\frac1{a^2}\right)\Cal{D}(a)\right]\notag\\
 &= \frac{\Delta_{\rm f}}{2A^2}\left(1-\frac{\Cal{D}(A)}{A}\right)\,,
 \label{eq:filex-t^fil(<R)}
 \end{align}
 so that we have 
 \be
 t^{\rm (fil)}(<R) = \frac13\langle{\Delta^{\rm (fil)}}\rangle(<R)\,. 
 \label{eq:t^fil=Delta^fil/3}
 \ee
 \emph{We expect this result to hold very generally}. To see why, consider that a Gaussian mixture (which we expect can accurately approximate any reasonable filamentary profile) would preserve this proportionality for each component and hence also for the total. As another direct example, a straightforward calculation shows that \eqn{eq:t^fil=Delta^fil/3} also holds for a power law profile $\Delta^{\rm (\rm fil)}(\mb{r})\propto (r_\perp/R_{\rm f})^{-\beta}$ with $0<\beta<2$, with $\langle{\Delta^{\rm (fil)}}\rangle(<R)\propto(R/R_{\rm f})^{-\beta}$ in this case. 

 Turning next to the halo self-contribution $\Delta^{\rm (halo)}(\mb{r})$ in \eqn{eq:filex-Delta}, we first note that our choice of spherical symmetry for the halo  means that the tidal term $t^{\rm (halo)}(<R)$ identically vanishes (section~\ref{subsubsec:sphsymm}), while the choice of the NFW profile means that $\Delta^{\rm (halo)}(<R)$ is given by the radial average of \eqn{eq:NFWgaussmix}.

 Figure~\ref{fig:filhalo} shows the halo-centric $\alpha(<4R_{\rm vir})$ in this model as a function of the ratio $R_{\rm vir}/R_{\rm f}$, setting the the filamentary relative overdensity $f_{\rm fil}=0.02\,(0.04)$ in the \emph{left (right) panel}. We have used values of $f_{\rm fil}$ and $R_{\rm f}$ approximately consistent with the distribution shown in Figure~39 of \citet{cvdwj14}; typical filaments at low redshift are expected to have overdensities $\avg{\Delta}\sim10$ and thicknesses of order $\sim\textrm{few}\times\Mpch$ \citep[see also][]{aragoncalvo+10,bprg17,kraljic+19,ftb19}.

 The solid curves show results when including the halo self-contribution as above.  There are two interesting features worth discussing.
 First, we clearly see that large haloes in thin filaments experience a weaker anisotropy: the curves decline at large $R_{\rm vir}/R_{\rm f}$. On the other hand, the anisotropy approaches a constant when $R_{\rm vir}\ll R_{\rm f}$. The value of this constant depends on $f_{\rm fil}$, with denser filaments (large $f_{\rm fil}$) producing a larger anisotropy. 

 More interestingly, we see that all the curves trace out nearly universal loci of $R_{\rm vir}/R_{\rm f}$ at fixed $f_{\rm fil}$, as expected from \eqns{eq:filex-Delta^fil(<R)} and~\eqref{eq:filex-t^fil(<R)} when the halo self-contribution is sub-dominant. 
 Due to the relation \eqref{eq:t^fil=Delta^fil/3} and the expectation that any filamentary profile can be approximated by a Gaussian mixture, we expect this near-universality to be very generally valid, with $R_{\rm f}$ being replaced by some characteristic scale describing the filament profile. 
 The dashed line in each panel of Figure~\ref{fig:filhalo} shows the value $1/2$ which would be obtained using \eqns{eq:t^fil=Delta^fil/3} and~\eqref{eq:alpha(<R):summary} assuming the halo contribution to be completely sub-dominant.
 This serves as an upper limit to the solid curves when the smoothing radius, $f_{\rm fil}$ or $R_{\rm f}$ are varied. 

 This behaviour is particularly interesting considering the fact that the value $\alpha(<4R_{\rm vir})\simeq0.5$ is known to cleanly segregate haloes living in filamentary environments from those in nodes, when filaments and nodes are defined using a counting of negative eigenvalues of the tidal tensor \citep[see, e.g., Figure 7 of][]{phs18a}. As mentioned above, we expect most realistic haloes in filaments to be found off-axis (and hence with substantially larger anisotropy than the solid curves in the Figure), with a small fraction being close to the axis. Our simple toy model then provides the first analytical explanation of why $\alpha(<4R_{\rm vir})\simeq0.5$ is a good segregator of filamentary environments: this would naturally arise if a large fraction of filamentary haloes are substantially off-axis \emph{and} sub-dominant in their contribution to $\avg{\Delta}(<4R)$. It will be very interesting to test these ideas by studying $\alpha(<4R_{\rm vir})$ as a function of the distance from the nearest filament \citep[e.g., using filament definitions such as the one in][]{sousbie11}, which we leave to future work.

 \subsection{Satellite in a spherical host}
 \label{subsec:satellite}
\begin{figure}
 \centering
 \includegraphics[width=0.45\textwidth]{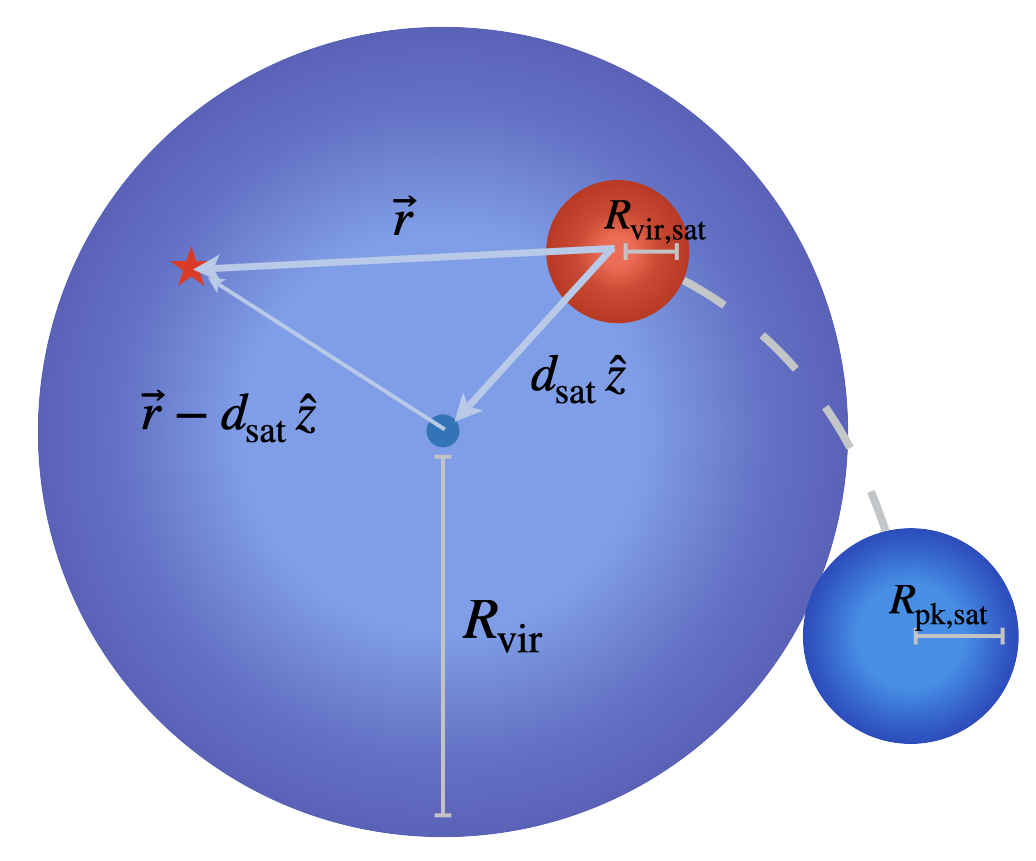}
 \caption{Illustration of a satellite of current radius $R_{\rm vir,sat}$ located at a distance $d_{\rm sat}$ from the centre of its host halo which has a spherically symmetric NFW density profile with radius $R_{\rm vir}$ and concentration $c_{\rm vir}$. The satellite is assumed to have started with a spherical NFW profile of radius $R_{\rm pk,sat}$ and concentration $c_{\rm pk,sat}$, which is assumed to be truncated due to mass loss as described in the text. From the vantage point of the satellite center, the spherical profile of the host is seen as having an axisymmetric anisotropy.}
 \label{fig:cartoon-satellite}
 \end{figure}

 \noindent
 For our next example, consider a spherically symmetric satellite of radius $R_{\rm sat,vir}$ located at distance $d_{\rm sat}$ from the centre of its spherically symmetric, NFW distributed host of radius $R_{\rm vir}$ and concentration $c_{\rm vir}$ (and hence a scale radius $r_{\rm s}$ as in equation~\ref{eq:rsrvircvir}). Figure~\ref{fig:cartoon-satellite} illustrates the situation.

 The dark matter overdensity in the vicinity of the satellite has two contributions, one from the satellite's own profile and the other from the profile of the host. We write the overdensity at position \mb{r} \emph{as measured from the satellite's location} as
 \be
 \Delta(\mb{r}) = \Delta^{\rm (host)}(\mb{r}) + \Delta^{\rm (sat)}(\mb{r})\,,
 \label{eq:satex-Delta}
 \ee
 with the superscripts on each term on the right indicating the two contributions.

 \subsubsection{Host contribution}
 \label{subsubsec:hostcontrib}
 \noindent
 Aligning the $z$-axis with the location of the halo centre as seen by the satellite and using the Gaussian mixture \eqref{eq:NFWgaussmix}, it is easy to show that the host contribution is given by the axisymmetric form
 \begin{align}
 \Delta^{\rm (host)}(\mb{r}) 
 &= \Delta_{\rm NFW}(|\mb{r}-d_{\rm sat}\hat z| \mid R_{\rm vir},c_{\rm vir}) \notag\\
 &= \sum_j\,w_j\,\Delta_{\rm G}(|\mb{r}-d_{\rm sat}\hat z| \mid\sigma_j,r_{\rm s})\notag\\
 &= \sum_j\,w_j\,\Delta_j\exp\left(-\frac{(r^2+d_{\rm sat}^2-2\,r\,d_{\rm sat}\mu)}{2\sigma_j^2r_{\rm s}^2}\right)\,.
 \label{eq:satex-Delta^host}
 \end{align}
 The corresponding multipole moments satisfy
 \begin{align}
 &\frac1{\sqrt{4\pi}}\Delta_{\ell0}^{\rm (host)}(r) + \delta_{\ell,0} \notag\\
 &\ph{\Delta}
 = \sqrt{2\ell+1}\,\sum_j\,w_j\,\Delta_j\,\e{-(r^2+d_{\rm sat}^2)/(2\sigma_j^2r_{\rm s}^2)} \notag\\
 &\ph{\Delta= \sqrt{2\ell+1}\,\sum_j}
 \times \int_{-1}^1\frac{\der\mu}{2}\,P_\ell(\mu)\,\e{\mu\,r\,d_{\rm sat}/(\sigma_j^2r_{\rm s}^2)}
 \label{eq:satex-Delta_ell0^host}
 \end{align}
 Using $P_0(\mu)=1$ and $P_2(\mu)=(3\mu^2-1)/2$, the integrals over $\mu$ are straightforward for $\ell=0$ and $\ell=2$. Defining 
 \be
 B_j \equiv d_{\rm sat}^2/(\sigma_j^2r_{\rm s}^2)\,,
 \label{eq:satex-Bdef}
 \ee
 \eqns{eq:t(<R)} and~\eqref{eq:delta(<R)} can be brought to the form
 \begin{align}
 &\langle{\Delta^{\rm (host)}}\rangle(<R) \notag\\
 &\ph{\Delta}
 = \sum_j\,w_j\,\Delta_j\,\e{-\frac12B_j}\left(\frac{d_{\rm sat}}{RB_j}\right)^3 \notag\\
 &\ph{\Delta\sum_jw_j}
 \times\int_0^{RB_j/d_{\rm sat}}\der a\,a\,\e{-a^2/(2B_j)}\sinh(a)\,,
 \label{eq:satex-Delta^host(<R)}\\
 &t^{\rm (host)}(<R) \notag\\
 &\ph{\Delta}
 = 2\sum_j\,w_j\,\Delta_j\,\e{-\frac12B_j}\left(\frac{d_{\rm sat}}{RB_j}\right)^3 \notag\\
 &\ph{2\Delta\sum_jw_j}
 \times\int_{RB_j/d_{\rm sat}}^\infty\frac{\der a}{a^4}\,\e{-a^2/(2B_j)}\notag\\
 &\ph{2\Delta\sum_jw_j\times\int_{RB_j}^\infty}
 \times\bigg[\left(a^2+3\right)\sinh(a)-3a\cosh(a)\bigg]\,.
 \label{eq:satex-t^host(<R)}
 \end{align}
 The integral over the auxiliary variable $a\propto r$ in \eqn{eq:satex-Delta^host(<R)} has a lengthy closed form expression in terms of error functions, which we omit for brevity. The integral in \eqn{eq:satex-t^host(<R)} must be performed numerically (although an asymptotic expansion, which we also omit here, is possible when $d_{\rm sat}\ll R$).

 \begin{figure}
 \centering
 \includegraphics[width=0.45\textwidth]{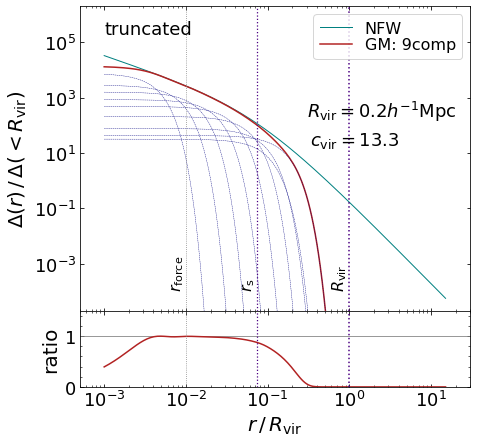}
 \caption{Example of truncated Gaussian mixture. Same as left panel of Figure~\ref{fig:gaussmixnfw}, discarding components having $\sigma_j>c_{\rm vir}/10$. This is useful in modelling the mass loss experienced by a satellite in a host halo. See text for a discussion.}
 \label{fig:gaussmixnfw-truncated}
 \end{figure}

 \subsubsection{Satellite contribution}
 \label{subsubsec:satcontrib}
 \noindent
 Turning next to the satellite self-contribution $\Delta^{\rm (sat)}(\mb{r})$ in \eqn{eq:satex-Delta}, as in section~\ref{subsec:filhalo} we first note that the tidal term $t^{\rm (sat)}(<R)$ identically vanishes for our spherical satellite, and the contribution $\langle{\Delta^{\rm (sat)}}\rangle(<R)$ is straightforward to compute once we decide on a model for the instantaneous satellite profile.

 Due to the collisionless nature of their dark matter content, satellite haloes orbiting their host rapidly lose mass due to processes such as tidal stripping and dynamical friction. While a detailed model of the resulting profile requires substantial numerical effort \citep[see, e.g.,][]{vdbg18,ogiya+19,svdb19}, our Gaussian mixtures approach suggests a simple and physically intuitive approximation; namely, we can \emph{mimic satellite mass loss by simply discarding appropriately chosen Gaussian components in the outskirts} of the satellite structure (like peeling off layers of mass one at a time). Figure~\ref{fig:gaussmixnfw-truncated} shows an example of an NFW Gaussian mixture truncated by discarding components having $\sigma_j > c_{\rm vir}/10$.\footnote{Alternatively, one could also build a separate, more detailed Gaussian mixture to accurately describe the shape of the truncated profile, rather than assuming the shape of the single Gaussian having the largest width of those remaining from the original mixture. We leave this to future work.}

 \begin{figure*}
 \centering
 \includegraphics[width=0.42\textwidth]{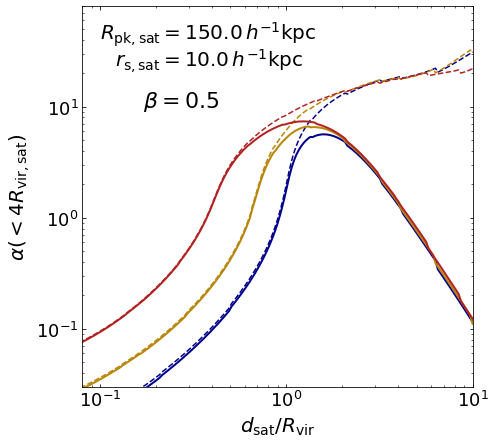}
 \includegraphics[width=0.42\textwidth]{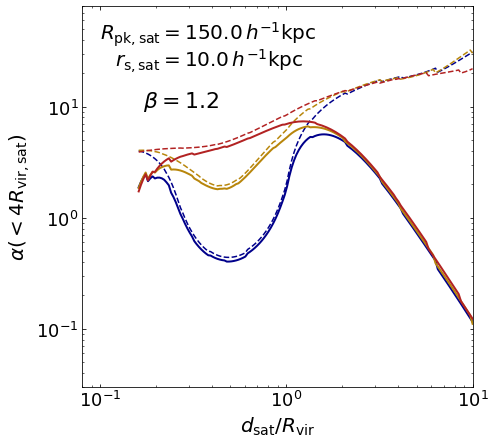}
 \vskip 0.025in
 \includegraphics[width=0.42\textwidth]{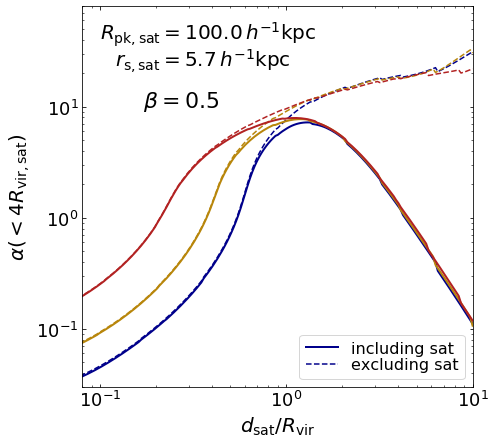}
 \includegraphics[width=0.42\textwidth]{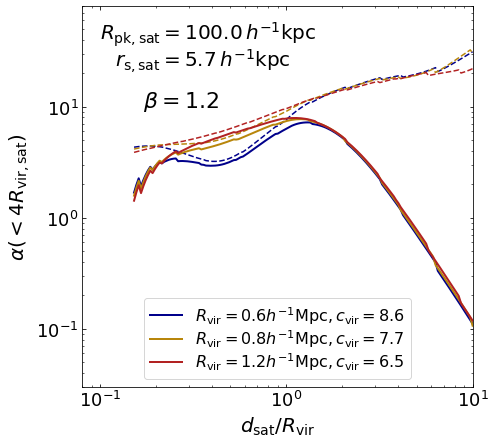}
 \caption{Tidal anisotropy $\alpha(<4R_{\rm vir,sat})$ centred on a satellite of current radius $R_{\rm vir,sat}$ at distance $d_{\rm sat}$ from the centre of its spherical NFW host of radius $R_{\rm vir}$ and concentration $c_{\rm vir}$. See Figure~\ref{fig:cartoon-satellite} for an illustration. Curves of different colours represent different combinations of $R_{\rm vir}$ and $c_{\rm vir}$ (with $R_{\rm vir}$ increasing from bottom to top around $d_{\rm sat}\sim R_{\rm vir}/2$ in all panels). Solid curves show results including the satellite profile which is modelled as a spherical NFW of radius $R_{\rm pk,sat}$ and scale radius $r_{\rm s,sat}$ (or concentration $c_{\rm pk,sat}=R_{\rm pk,sat}/r_{\rm s,sat}$) as indicated, and truncated at the current radius $R_{\rm vir,sat}$ as decribed in the text. \emph{Upper} and \emph{lower panels} show results for two choices of combinations of $R_{\rm pk,sat}$ and $r_{\rm s,sat}$, as indicated in the plot labels. $R_{\rm vir,sat}$ is calculated using a toy model of mass loss (equation~\ref{eq:satex-Rsatmodel}) with $\beta=0.5$ in the \emph{left panels} and $\beta=1.2$ in the \emph{right panels}. Dashed curves show results excluding the satellite profile (i.e., assuming the satellite contributes negligible density). All curves truncate at small distances where $R_{\rm vir,sat} < r_{\rm s,sat}$.}
 \label{fig:satellite}
 \end{figure*}

 More precisely, let the satellite start with an NFW profile at infall, with virial radius $R_{\rm pk,sat}$ and concentration $c_{\rm pk,sat}=R_{\rm pk,sat}/r_{\rm s,sat}$, with a corresponding Gaussian mixture $\{w_k^{\rm (sat)},\sigma_k^{\rm (sat)}\}$ (which is obviously distinct from the host's mixture $\{w_j,\sigma_j\}$). If the current radius of the satellite is $R_{\rm vir,sat}$, then we model its current profile by discarding the Gaussian components having $\sigma_k^{\rm (sat)} > R_{\rm vir,sat}/r_{\rm s,sat}$ (recall the widths are in units of the NFW scale radius). We also assume that the satellite will be quickly disrupted if its radius becomes smaller than its initial scale radius $r_{\rm s,sat}$, so we do not calculate $\alpha$ in this regime.

 It remains to decide the current radius $R_{\rm vir,sat}$, which is determined by the mass loss rate integrated over the satellite orbit. Typically, mass loss for satellites is close to being exponential in the number of dynamical times elapsed since first infall into the host \citep[see, e.g.,][]{vdbtg05}. Radial infall would suggest a scaling $\Delta t\propto (R_{\rm vir}-d_{\rm sat})^{3/2}$ for the time $\Delta t$ spent by the satellite inside the host before first pericentre passage. Initial angular momentum would modify this scaling and allow more time to be spent at larger separations from the host centre. Finally, if the satellite is outside the host ($d_{\rm sat} > R_{\rm vir}$), we assume there is no mass loss.\footnote{We are also ignoring mass accretion for both satellite and host from sources other than their interaction, which would require a more detailed model of the external environment of the host.} We therefore explore the heuristic model
 \be
 \frac{R_{\rm vir,sat}}{R_{\rm pk,sat}} = {\rm min}\left\{1,\exp\left[\frac13\left(1-(R_{\rm vir}/d_{\rm sat})^\beta\right)\right]\right\}
 \label{eq:satex-Rsatmodel}
 \ee
 parametrised by $\beta$. 
 The resulting $\langle{\Delta^{\rm (sat)}}\rangle(<R)$ adds to the halo contribution from \eqn{eq:satex-Delta^host(<R)} to give $\avg{\Delta}(<R)$, while $t(<R)=t^{\rm (host)}(<R)$ from \eqn{eq:satex-t^host(<R)}, as we argued above. 

 \subsubsection{Results}
 \label{subsubsec:satellite-results}
 \noindent
 We are interested in the resulting behaviour of $\alpha(<4R_{\rm vir,sat})$. Figure~\ref{fig:satellite} shows some examples using $\beta=0.5$ and $\beta=1.2$ (\emph{left} and \emph{right columns}, respectively) and large/small values of $R_{\rm pk,sat}$ and $r_{\rm s,sat}$ (\emph{top}/\emph{bottom} rows, respectively). The solid and dashed curves show results when including or excluding, respectively, the satellite self-contribution. (For the latter, we omit adding the satellite overdensity, similarly to omitting the halo contribution in section~\ref{subsec:filhalo}). 

 Several interesting features are apparent in the plots. First, the satellite self-contribution plays a dominant role in decreasing the tidal anisotropy in two regimes, both far outside the host (in all cases) and deep in its inner region (in the $\beta=1.2$ model, just before the mass loss makes $R_{\rm vir,sat}<r_{\rm s,sat}$). The behaviour outside the host is easy to understand: in this regime, the satellite has a fixed NFW profile with $R_{\rm vir,sat}=R_{\rm pk,sat}$ in our model, while the contribution of the host to both $t(<4R_{\rm pk,sat})$  and $\avg{\Delta}(<4R_{\rm pk,sat})$ becomes increasingly small at larger $d_{\rm sat}$. As a result, the satellite sees an increasingly isotropic environment at larger separations from the host, since we have not included any effect of the cosmic web in this model. The effect in the inner region, on the other hand, is driven by the rapid decrease of $R_{\rm vir,sat}$ at small $d_{\rm sat}$ (equation~\ref{eq:satex-Rsatmodel}) which leads to a rapid increase of $\langle\Delta^{(\rm sat)}\rangle(<4R_{\rm vir,sat})$.\footnote{The jaggedness apparent in the solid curves in the right panels of Figure~\ref{fig:satellite} just before the curves truncate at the left, is due to the successive discrete removal of Gaussian components from the satellite profile to mimic mass loss. The jaggedness in the dashed curves at large $d_{\rm sat}$, on the other hand, is due to numerical artefacts in the integral in \eqn{eq:satex-t^host(<R)}, which converges slowly for Gaussian components with large widths.} 
 The corresponding increase in the host contribution $\langle\Delta^{(\rm host)}\rangle(<4R_{\rm vir,sat})$ is a weaker function of $d_{\rm sat}$ since the local density due to the host is a relatively smooth function in the satellite vicinity. This effect occurs at larger $d_{\rm sat}$ in the $\beta=1.2$ model in which the decrease of $R_{\rm vir,sat}$ is faster. In both of these regimes, we can also see that there is essentially no dependence on the virial radius and scale radius of the host.

 Secondly, at separations $d_{\rm sat}\lesssim R_{\rm vir}$ we see a drop in the anisotropy strength whose magnitude is evidently a strong function of the relative size of the satellite to the host, but is \emph{essentially independent of the satellite's self-contribution}. 
 The latter observation suggests that this effect is entirely driven by the fact that the host NFW profile is being smoothed with different smoothing scales $R_{\rm smooth}=4R_{\rm sat,vir}$ at different $d_{\rm sat}$ and for different model parameters. 
 Larger values of $\lambda\equiv R_{\rm smooth}/R_{\rm vir}$ at the same $d_{\rm sat}/R_{\rm vir}$ produce a lower anisotropy, which is sensible (in the limit $\lambda\gg d_{\rm sat}/R_{\rm vir}$, the anisotropy should become negligible).
 Indeed, the strongest decrease is seen in the \emph{top left} panel (larger $R_{\rm pk,sat}$, weaker mass loss rate) for the smallest host size (blue curves), all of which conspire to produce the largest $\lambda$ at fixed $d_{\rm sat}/R_{\rm vir} \lesssim 1$. As the host size increases (red and yellow curves in same panel), or as the initial satellite radius decreases (bottom panels), or as the mass loss rate increases (right panels), $\alpha$ decreases by smaller amounts. 

 For a fixed host size, the $\beta=1.2$ model shows a minimum in $\alpha$ for the smaller host sizes. The fact that $\alpha$ increases as the separation decreases below $d_{\rm sat}\lesssim R_{\rm vir}/2$ in all but the largest hosts, can be understood as an interplay of two effects: the smoothing radius is decreasing and the smoothing centre is moving closer to the centre of symmetry. The latter effect is initially dominant (at $d_{\rm sat}\lesssim R_{\rm vir}$) where $\lambda$ is large. At smaller $d_{\rm sat}$, the decrease in $\lambda$ becomes more rapid (equation~\ref{eq:satex-Rsatmodel}), so that moving closer to the symmetry centre plays a weaker role than the fact that the anisotropic tidal field is being smoothed by a lesser amount. This leads to a rise in $\alpha$ beyond the point where these effects balance. The effect is not seen in the $\beta=0.5$ cases over the range of scales plotted, consistent with the weaker mass loss in this model. Since $d_{\rm sat}\to0$ would bring us exactly to the symmetry center, $\alpha$ is expected to \emph{decrease} again in the $\beta=1.2$ case for smaller $d_{\rm sat}$ (provided the smoothing radius remains non-zero), even when ignoring the satellite self-contribution. Indeed, there are indications of a maximum in $\alpha$ in the curves excluding the satellite, which coincidentally occur just before $R_{\rm vir,sat}$ becomes smaller than $r_{\rm s,sat}$.

 Finally, the combined result of all these effects is to produce an \emph{overall maximum} in $\alpha(<4R_{\rm vir,sat})$ at $d_{\rm sat}\gtrsim R_{\rm vir}$ just outside the host, with maximum values between $5$-$8$. This is in sharp contrast to the statistics of subhaloes seen in cosmological simulations which show typical values of $\alpha(<4R_{\rm sat,vir})\lesssim1.0$ \citep{zphp20}. This is almost certainly a consequence of the fact that most satellites identified in a given simulation snapshot are \emph{not} close to their first infall; satellites found further inside their host are indeed predicted to have smaller $\alpha$ in our models.

 This suggests a further interesting application of our formalism. For a satellite population with some typical mass ratio $\mu_{\rm sat}\sim(R_{\rm vir,sat}/R_{\rm sat})^3$ and typical separation $d_{\rm sat}/R_{\rm vir}$, measurements of $\alpha(<4R_{\rm vir,sat})$ in simulations could in principle be used to constrain empirical models of mass loss by comparing with plots such as Figure~\ref{fig:satellite}. This is an exercise we will pursue in future work.

 \subsection{Anisotropic halo}
 \label{subsec:anisohalo}
 \noindent
 As our final example, we consider the tidal field of an axisymmetric halo. Simulated haloes are well-known to be triaxial \citep[see, e.g.][]{allgood+06}, with shapes that correlate with their large-scale environment \citep{fw10}. The triaxiality of haloes is also known to be well-described using anisotropically scaled NFW profiles \citep{js02}. It is therefore interesting to explore the tidal influence of the anisotropic halo shape. 

 However, any such model involving smoothing at scales larger than the virial radius must account for the tidal influence of the halo \emph{environment} as well, as we approximated in each of the models above. For the situation we are now interested in, we would need to also model genuine 2-halo effects due to large-scale structure \citep{ps13}.\footnote{We can always consider an axisymmetric halo at the axis of and perfectly aligned with a filament. The resulting tidal anisotropy would be a straightforward combination of the results of this section and those of section~\ref{subsec:filhalo}. Apart from this somewhat contrived example, modelling the intermediate-scale environment of an anisotropic halo is more challenging.} A model focusing only on the halo profile, on its own, cannot provide realistic insight into the behaviour of the halo-centric $\alpha(<4R_{\rm vir})$. In the interest of completeness, however, we will show the results for the 1-halo contribution of such a model, leaving a fuller exploration of effects beyond the halo radius to future work.

 With these caveats in mind, consider an axisymmetric halo profile given by
 \begin{align}
 \Delta(\mb{r}) &= \Delta_{\rm NFW}\left(\sqrt{z^2+(x^2 + y^2)/b^2}|R_{\rm vir},c_{\rm vir}\right) \notag\\
 &= \Delta_{\rm NFW}\left(\frac{r}{b}\,\sqrt{1-\left(1-b^2\right)\mu^2}|R_{\rm vir},c_{\rm vir}\right)\,,
 \label{eq:haloex-Delta}
 \end{align}
 where $0<b\leq1$ is the common value of the intermediate-to-major and minor-to-major eigenvalue ratios of the halo moment-of-inertia tensor (with $b=1$ describing a spherical halo) and we aligned the major axis with the Cartesian $z$-axis.

 The resulting calculation is in fact very similar to the one in section~\ref{subsec:filhalo}, since the anisotropy in the density has a similar (but not the same) structure as the filamentary profile studied there. Applying the Gaussian mixture \eqref{eq:NFWgaussmix}, the $\ell=0,2$ multipole moments (equation~\ref{eq:Delta-ell0}) of each individual Gaussian component can be written as
 \begin{align}
 \Delta_{00}(r|\sigma_j,r_{\rm s}) &=\sqrt{4\pi}\left(\Delta_{\rm G}(r|\sigma_j,r_{\rm s})\,\frac{\Cal{D}\left(\kappa_j\right)}{\kappa_j}-1\right)\,,\notag\\
 \Delta_{20}(r|\sigma_j,r_{\rm s}) &=\sqrt{5\pi}\,\Delta_{\rm G}(r|\sigma_j,r_{\rm s}) \notag\\
 &\ph{\sqrt{5}}
\times\left[\frac3{2\kappa_j^2}-\frac{\Cal{D}\left(\kappa_j\right)}{\kappa_j}\left(1+\frac3{2\kappa_j^2}\right)\right]\,,
 \label{eq:haloex-Delta_ell0}
 \end{align}
 where $\Delta_{\rm G}(r|\sigma_j,r_{\rm s})$ was defined in \eqn{eq:singlegauss}, Dawson's integral $\Cal{D}(z)$ was defined in \eqn{eq:Dawsondef}
 and we defined
 \be
 \kappa_j(r) \equiv \frac{\sqrt{1-b^2}}{\sqrt{2}\,b}\left(\frac{r}{r_{\rm s}\sigma_j}\right)\,.
 \label{eq:kappajdef}
 \ee
 Equations~\eqref{eq:haloex-Delta_ell0} can be integrated (see equations~\ref{eq:t(<R)} and~\ref{eq:delta(<R)}) to give the single-component contributions to $\avg{\Delta}(<R)$ and $t(<R)$:
 \begin{align}
 &\avg{\Delta}(<R|\sigma_j,r_{\rm s}) \notag\\
 &= {3\Delta_jb^2\left(\frac{r_{\rm s}\sigma_j}{R}\right)^3\bigg[\sqrt{\frac\pi2}\,\erf{\frac{R}{\sqrt{2}r_{\rm s}\sigma_j}}} \notag\\
 &\ph{3\Delta_jb^2\left(\frac{r_{\rm s}\sigma_j}{R}\right)^3\bigg[\bigg]}
 {- \frac{\sqrt{2}\,b}{\sqrt{1-b^2}}\,\e{-R^2/(2r_{\rm s}^2\sigma_j^2)}\,\Cal{D}\left(\kappa_j(R)\right)\bigg]}\notag\\
 &t(<R|\sigma_j,r_{\rm s}) \notag\\
 &= \frac{3\Delta_j}2\,\int_{\kappa_j(R)}^\infty\frac{\der s}{s^2}\,\e{-b^2s^2/(1-b^2)}\,\left[\frac1s-\left(\frac23+\frac1{s^2}\right)\,\Cal{D}(s)\right]\,.
 \label{eq:haloex-Delta_ell0-integ}
 \end{align}
 The integral defining $t(<R|\sigma_j,r_{\rm s})$ must be performed numerically in this case, but is straightforward to compute. 

 \begin{figure}
 \centering
 \includegraphics[width=0.475\textwidth]{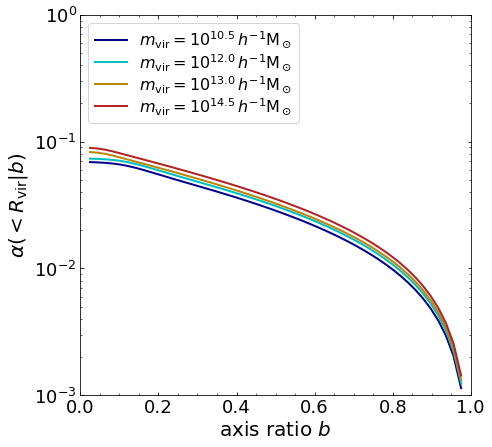}
 \caption{Halo-scale tidal anisotropy $\alpha(<R_{\rm vir})$ for an axisymmetric NFW halo with profile \eqref{eq:haloex-Delta}, as a function of axis ratio $b$ for a range of values of virial mass $m_{\rm vir}$ (coloured curves, decreasing in $m_{\rm vir}$ from bottom to top). We see a sharp increase in $\alpha$ as $b$ decreases from unity and the haloes become aspherical, followed by a slower increase to a finite value as $b\to0$. The anisotropy depends only weakly on halo mass.}
 \label{fig:anisohalo}
 \end{figure}

 Figure~\ref{fig:anisohalo} shows the resulting total $\alpha(<R_{\rm vir})$ (note the smoothing scale) for a few halo masses $m_{\rm vir}$ as a function of axisymmetry parameter $b$. We see little dependence on $m_{\rm vir}$ for all $b$. At fixed $m_{\rm vir}$, on the other hand, $\alpha(<R_{\rm vir})$ sharply increases as $b$ decreases from unity and approaches a value $\lesssim0.1$ as $b\to0$. The trend as $b\to1$ is sensible, since the anisotropy must vanish for a perfectly spherical halo. As we mentioned above, however, it is difficult to draw substantial further insight from this model due to the lack of environmental contributions to $\alpha$.

 \section{Summary \& Conclusion}
 \label{sec:conclude}
 \noindent
Motivated by the need to understand the nature of the tidal field of different cosmic web environments, in this work we have presented a descriptive analytical framework for studying the primary tool of interest, namely, the halo-centric tidal tensor spherically averaged on a smoothing scale proportional to the halo virial radius $R_{\rm vir}$. Specifically, we were interested in studying the `tidal anisotropy' scalar $\alpha(<4R_{\rm vir})$ (equation~\ref{eq:alphadef}) in different cosmic web environments. 

Although the formalism we developed in section~\ref{sec:TT} and Appendix~\ref{app:analytical} is capable of describing arbitrarily anisotropic tidal environments, we focused on axisymmetric anisotropies in this work, with the intent of building the simplest possible examples capable of providing insight into the behaviour of $\alpha(<4R_{\rm vir})$. 
Equations~\eqref{eq:t(<R)}, \eqref{eq:delta(<R)} and~\eqref{eq:alpha(<R):summary}, valid for any axisymmetric density distribution \eqref{eq:Delta-genericaxisymm}, form the core result of this work.

As a simple application of this formalism, we proved in general that any misalignment between the spherically averaged versions of the tidal tensor \eqref{eq:TTcoordinvar} and density Hessian \eqref{eq:hessiandef} \emph{must be a consequence of departures from axisymmetry in the unsmoothed density}. In other words, the misalignment angle between these two \emph{smoothed} tensors can be thought of as a direct measurement of triaxiality of the \emph{unsmoothed} field. This is an interesting example of a situation where smoothing with a spherical filter does not erase anisotropic information.

As examples of interesting cosmological situations, we studied three toy models (section~\ref{sec:applicn}): 
 \begin{itemize}
 \item a spherical halo at the axis of a cylindrical filament, 
 \item an off-centred satellite in a spherical host halo, and 
 \item a halo with an axisymmetric density profile,
 \end{itemize}
 each of which results in a single special direction that serves as the axis of symmetry (see Figures~\ref{fig:cartoon-filament} and~\ref{fig:cartoon-satellite}).

 In each of these cases, we used the spherical NFW profile (equation~\ref{eq:NFWprofile}) as a building block to describe halo density profiles and satellite environments. Despite its simplicity, the NFW profile typically does not admit closed form expressions for the integrals defining the spherically averaged tidal tensor components, even for the restricted axisymmetric case. \emph{A key simplification occurs by approximating the NFW profile itself as a mixture of spherical Gaussians} (section~\ref{sec:spherical}, equation~\ref{eq:NFWgaussmix}). Although this converts a 2-parameter function into a sum of multiple ($\sim20$) Gaussian components, the resulting ability to analytically calculate $\alpha(<4R_{\rm vir})$ for each of our examples is, we believe, worth the additional complexity. We used standard numerical libraries to implement a non-negative least squares fit of the Gaussian mixture to the NFW profile (Figures~\ref{fig:gaussmixnfw} and~\ref{fig:error-gmNFW}), which was then propagated into the formalism for the tidal tensor.

 The Gaussian mixture lends itself to considerable flexibility, as we showed in our toy models. For example, in the case of the halo in a filament (section~\ref{subsec:filhalo}), in addition to the halo's NFW profile, the \emph{filament's} density profile can also be modelled as a Gaussian mixture (we used only one component for illustration). We showed that the resulting tidal tensor has a remarkably simple, closed form analytical structure (equation~\ref{eq:t^fil=Delta^fil/3}) such that $\alpha^{\rm (fil)}(<R) = 1/2$ when ignoring the halo's self-contribution, for any smoothing scale. While the more realistic calculation including the halo contribution changes this to the behaviour seen in Figure~\ref{fig:filhalo}, we argued that our calculation provides the first analytical insight into the observed fact that filamentary haloes in simulations are bounded by $\alpha(<4R_{\rm vir})\gtrsim0.5$ \citep[see, e.g., Figure 7 of][]{phs18a}. 

 For the satellite of radius $R_{\rm vir,sat}$ off-centred from its host (section~\ref{subsec:satellite}), the Gaussian mixture describing the satellite allowed us to easily approximate mass loss due to various dynamical mechanisms by simply discarding the Gaussian components in the satellite's outskirts (illustrated in Figure~\ref{fig:gaussmixnfw-truncated}). What constitutes the outskirts is decided by the specific model of mass loss; we argued that our formalism can be adapted in a straightforward manner to constrain such models by comparing predictions such as those in Figure~\ref{fig:satellite} with measurements of $\alpha(<4R_{\rm vir,sat})$ in simulations. The simplified mass-loss models we explored also generically predict that $\alpha(<4R_{\rm vir,sat})$ has its maximum close to first infall, which should be possible to test in simulations.

 Finally, we included the example of the axisymmetric halo for completeness (section~\ref{subsec:anisohalo}, Figure~\ref{fig:anisohalo}), noting that a fuller understanding of $\alpha(<4R_{\rm vir})$ for such objects would require the inclusion of models of their external cosmic environment \citep[e.g., along the lines described by][]{ps13}.

 Our formalism and associated models allow for several immediate applications and extensions, apart from those mentioned above. We conclude by discussing some of these here.
 \begin{itemize}
 \item 
 Although the configurations we studied focused on filaments and substructure, haloes in voids are interesting too \citep[e.g.,][]{rieder+13}, and can be easily described by our axisymmetric formalism similarly to our off-centred satellite example.
 \item
 For haloes in filaments, as we argued in section~\ref{subsec:filhalo}, a systematic study in $N$-body simulations of $\alpha(<4R_{\rm vir})$ as a function of distance from the filament axis would further help clarify the significance of the value $\alpha(<4R_{\rm vir})=1/2$ for such objects \citep[see, e.g.,][who show that dynamical effects due to tidal fields can occur over distances of several tens of Mpc]{pvdwj08}. Moreover, the near-universality of $\alpha(<4R_{\rm vir})$ with filament size seen in Figure~\ref{fig:filhalo} (if it persists for a broader population than perfectly on-axis haloes) might be useful in constraining filamentary profile shapes in simulations, using only measurements of halo-centric $\alpha(<4R_{\rm vir})$.
 \item
 There has also been considerable discussion in the literature as to which variables defined at which smoothing scales are the most suitable indicators of halo properties and large-scale correlations, with the fixed-scale overdensity $\avg{\Delta}(<R)$ with $R\simeq1$-$2\Mpch$ and halo-scaled tidal anisotropy $\alpha(<4R_{\rm vir})$ being primary contenders \citep{yfw13,han+19,goh+19,rphs19}. Our analytical formalism would be a useful tool in disentangling some of these correlations, e.g., by studying $\alpha(<R)$ as a function of $\avg{\Delta}(<R)$ for different choices of $R$ and in different cosmic web environments.
 \item
 Finally, an extension of the formalism to include anisotropies beyond axisymmetry would allow us to study more realistic (sub)halo configurations such as off-axis haloes in a filament, off-centred satellites in an aspherical host, aspherical haloes in a void, etc. This extension is tedious, but feasible (see the discussion at the end of Appendix~\ref{app:axisymm}). The associated Gaussian mixtures description of halo profiles with substructure would also have potential applications in semi-analytic modelling for gravitational lensing studies, galaxy cluster modelling, etc. 
 \end{itemize}
 We will return to these problems in future studies.

 \section*{Acknowledgments}
 It is a pleasure to thank Ravi Sheth and Oliver Hahn for many insightful discussions, and the anonymous referee for a constructive report.
This research is supported by the Associateship Scheme of ICTP, Trieste and the Ramanujan Fellowship awarded by the Department of Science and Technology, Government of India. 
This work used the open source computing packages NumPy \citep{vanderwalt-numpy}\footnote{\href{http://www.numpy.org}{http://www.numpy.org}}, SciPy \citep{scipy}, Matplotlib \citep{hunter07_matplotlib}\footnote{\href{https://matplotlib.org/}{https://matplotlib.org/}} and Jupyter\footnote{\href{https://jupyter.org}{https://jupyter.org}} Notebook.

\section*{Data Availability}
No new data were analysed in support of this research. Python code for producing the Gaussian mixtures and analytical models used in this work, along with a Jupyter notebook containing example calculations that reproduce all the main plots of the paper, can be downloaded from \href{https://bitbucket.org/aparanjape/gaussmixnfw}{https://bitbucket.org/aparanjape/gaussmixnfw}.

\bibliography{masterRef}

\appendix

\section{Details of calculations}
\label{app:analytical}
\noindent
In this Appendix we build up the analytical formalism used in the text to calculate the spherically averaged tidal tensor and the scalar tidal anisotropy $\alpha(<R)$.

\subsection{Tidal tensor in polar coordinates}
\label{app:TTpol}
\noindent
We will use the usual spherical polar coordinates $\{r,\theta,\phi\}$ related to Cartesian coordinates $\{x,y,z\}$ through
\begin{align}
x &= r \,s_{\theta} \,c_{\phi} \quad;\quad y = r \,s_{\theta} \,s_{\phi} \quad;\quad z = r \,c_{\theta} 
\label{eq:pol-to-xyz}
\end{align}
As in the main text, we abbreviate $\sin(\theta)=s_\theta$ and $\cos(\theta)=c_\theta$ for any angle $\theta$. We will also use the local rotation relating the unit vectors in the Cartesian and spherical polar bases \citep{binney-tremaine-GalDyn}
\begin{align}
\hat r &= s_\theta\,\left(c_\phi\,\hat x + s_\phi\,\hat y\right) + c_\theta\,\hat z\notag\\
\hat \theta &= c_\theta\,\left(c_\phi\,\hat x + s_\phi\,\hat y\right) - s_\theta\,\hat z\notag\\
\hat \phi &= -s_\phi\,\hat x + c_\phi\,\hat y\,,
\label{eq:polunitvecs}
\end{align}
which can be summarised as 
\be
\mb{e}_\alpha = R^{\ph{\alpha}i}_\alpha\,\mb{e}_i \quad;\quad \mb{e}_i = R^{\alpha}_{\ph{\alpha}i}\,\mb{e}_\alpha\,,
\label{eq:unitvecrot}
\ee
where repeated indices are summed, with $\alpha=\{r,\theta,\phi\}$ and $i=\{x,y,z\}$, and where the local rotation matrix $R^{\ph{\alpha}i}_\alpha$ is given by
\begin{align}
R^{\ph{\alpha}i}_\alpha &= 
\begin{pmatrix}
s_\theta\, c_\phi & s_\theta\, s_\phi & c_\theta \\
c_\theta\, c_\phi & c_\theta\, s_\phi & -s_\theta \\
-s_\phi & c_\phi & 0
\end{pmatrix}\,,
\label{eq:rotmat}
\end{align}
with the inverse given by $R^\alpha_{\ph{\alpha}i} = (R^{\ph{\alpha}i}_\alpha)^{-1} =  (R^{\ph{\alpha}i}_\alpha)^{\rm T}$ (our convention is that the upper index is the column index and the lower index is the row index). 

For any vector $\mb{v} = v_i\,\mb{e}_i = v_\alpha\,\mb{e}_\alpha$, the transformations relating the Cartesian components $\{v_i\}$ and the spherical components $\{v_\alpha\}$ are
\be
v_i = R^\alpha_{\ph{\alpha}i}\,v_\alpha\quad;\quad v_\alpha  = R^{\ph{\alpha}i}_\alpha\,v_i\,.
\label{eq:vpol-to-vxyz}
\ee
Similarly, for any symmetric tensor $T = T_{ij}\left(\mb{e}_i\otimes\mb{e}_j\right) = T_{\alpha\beta}\left(\mb{e}_\alpha\otimes\mb{e}_\beta\right)$, we have 
\be
T_{ij} = R^\alpha_{\ph{\alpha}i}\,R^\beta_{\ph{\alpha}j}\,T_{\alpha\beta} \quad;\quad T_{\alpha\beta} = R^{\ph{\alpha}i}_\alpha\,R^{\ph{\alpha}j}_\beta\,T_{ij}\,.
\label{eq:tensorrot}
\ee
Explicitly, we have
\begin{align}
T_{xx} &= c_\phi^2\,A_+ + s_\phi^2\,T_{\phi\phi} - s_{2\phi}\,B_+\notag\\
T_{yy} &= s_\phi^2\,A_+ + c_\phi^2\,T_{\phi\phi} + s_{2\phi}\,B_+\notag\\
T_{zz} &= A_-\notag\\
T_{xy} &= \frac12s_{2\phi}\left(A_+-T_{\phi\phi}\right) + c_{2\phi}\,B_+\notag\\
T_{xz} &= c_\phi\,C - s_\phi\,B_-\notag\\
T_{yz} &= s_\phi\,C + c_\phi\,B_-\,,
\label{eq:Tpol-to-Txyz}
\end{align}
where we defined the quantities
\begin{align}
A_+ &\equiv s_\theta^2\,T_{rr} + c_\theta^2\,T_{\theta\theta} + s_{2\theta}\,T_{r\theta}\notag\\
A_- &\equiv c_\theta^2\,T_{rr} + s_\theta^2\,T_{\theta\theta} - s_{2\theta}\,T_{r\theta}\notag\\
B_+&\equiv s_\theta\,T_{r\phi} + c_\theta\,T_{\theta\phi}\notag\\
B_-&\equiv c_\theta\,T_{r\phi} - s_\theta\,T_{\theta\phi}\notag\\
C &\equiv \frac12 s_{2\theta}\left(T_{rr}-T_{\theta\theta}\right) + c_{2\theta}\,T_{r\theta}\,.
\label{eq:A+-B+-C}
\end{align}
As a check, a direct calculation shows that the linear and quadratic rotational invariants $I_1$ and $I_2$ obey the identities
\begin{align}
I_1 &= T_{xx} + T_{yy} + T_{zz} = T_{rr} + T_{\theta\theta} + T_{\phi\phi}\notag\\ 
I_2 &= T_{xx}T_{yy} - T_{xy}^2 + T_{xx}T_{zz} - T_{xz}^2 + T_{yy}T_{zz} - T_{yz}^2 \notag\\
     &=  T_{rr}T_{\theta\theta} - T_{r\theta}^2 + T_{rr}T_{\phi\phi} - T_{r\phi}^2 + T_{\theta\theta}T_{\phi\phi} - T_{\theta\phi}^2\,,
\label{eq:I1I2}
\end{align}
as expected from the fact that \eqns{eq:polunitvecs} constitute a rotation of basis vectors. Note that $q^2=I_1^2-3I_2$ (equation~\ref{eq:q2def}). These expressions will be useful in building $\alpha(<R)$ later. 

Consider the tidal tensor $T$ defined in \eqn{eq:TTcoordinvar}. The spherical polar components $T_{\alpha\beta}$ can be expressed in terms of spherical polar derivatives of $\psi$ using \eqn{eq:gradpol} and the derivative identities
\begin{align}
\p_\theta\,\hat r &= \hat\theta \quad;\quad \p_\theta\,\hat\theta = -\hat r \quad;\quad \p_\phi\,\hat\phi = -\left(s_\theta\,\hat r+c_\theta\,\hat\theta\right) \notag\\
\p_\phi\,\hat r &= s_\theta\,\hat\phi \quad;\quad \p_\phi\,\hat\theta = c_\theta\,\hat \phi\,,
\label{eq:polunitvecderivs}
\end{align}
with all other derivatives of the polar unit vectors being zero. 
This leads to \eqn{eq:TTpol}.
As a check, note that the trace of the tensor recovers the Laplacian of the potential in polar coordinates:
\begin{align}
T_{rr} &+ T_{\theta\theta} + T_{\phi\phi} \\
&= \p_r^2\psi + \frac2r\,\p_r\psi + \frac1{r^2}\left[\p_\theta^2\psi + \frac{c_\theta}{s_\theta}\,\p_\theta\psi + \frac1{s_\theta^2}\,\p_\phi^2\psi\right] \notag\\
&= \frac1{r^2}\p_r\left(r^2\,\p_r\psi\right) + \frac1{r^2}\left[ \frac1{s_\theta}\p_\theta\left(s_\theta\,\p_\theta\psi\right) + \frac1{s_\theta^2}\,\p_\phi^2\psi\right]\notag\\
&= \nabla^2\psi \,.
\label{eq:TTpoltrace}
\end{align}

\subsection{Multipole expansions}
\label{app:multipole}
\noindent
Here we collect some useful standard results concerning spherical harmonics and related functions. The spherical harmonics are given by
\be
Y^m_\ell(\hat r) = N_{\ell m}\,P^m_\ell(\mu)\,\e{im\phi}\,,
\label{eq:Ylm}
\ee
with $i=\sqrt{-1}$ and $P^m_\ell(\mu)$ being the associated Legendre functions satisfying
\be
P^m_\ell(\mu) = \frac{(-1)^m}{2^\ell\ell!}\,\left(1-\mu^2\right)^{m/2}\frac{\der^{\ell+m}}{\der\mu^{\ell+m}}\left(\mu^2-1\right)^\ell,
\label{eq:Pml}
\ee
and where $N_{\ell m}$ is a normalisation constant given by
\be
N_{\ell m} = \sqrt{\frac{(2\ell+1)}{4\pi}\,\frac{(\ell-m)!}{(\ell+m)!}}\,.
\label{eq:Ylm-norm}
\ee
The following identities are useful:
\begin{align}
\frac1{s_\theta}\,\p_\theta f &= -\p_\mu f \notag\\ 
\p_\theta^2 f &= -\mu\p_\mu f + (1-\mu^2)\p_\mu^2f \notag\\
(\mu^2-1)\p_\mu P_\ell &= \ell\left(\mu\,P_\ell(\mu)-P_{\ell-1}(\mu)\right)\notag\\
\p_\theta^2P_\ell &= -\ell(\ell+1)P_\ell + \mu\p_\mu P_\ell\notag\\
s_{2\theta}\p_\theta P_\ell &= 2\ell\mu\left(\mu P_\ell-P_{\ell-1}\right)\,,
\label{eq:thetaderivs}
\end{align}
where $f$ is any function of $\mu$ and $P_\ell(\mu) = P^{m=0}_\ell(\mu)$ (see equation~\ref{eq:Pml}) are Legendre polynomials satisfying the integral relations
\begin{align}
\int_{-1}^1\frac{\der\mu}{2}\,P_\ell(\mu) &= \delta_{\ell,0}\,,
\label{eq:Legendreint}\\
\int_{-1}^1\frac{\der\mu}{2}\,P_\ell(\mu)\,P_{\ell^\prime}(\mu) &= \frac1{2\ell+1}\,\delta_{\ell,\ell^\prime}\,,
\label{eq:Legendreortho}
\end{align}
where $\delta_{\ell,\ell^\prime}$ is the Kronecker delta symbol.

\subsection{Azimuthal averaging}
\label{app:azavg}
As discussed in section~\ref{subsec:sphavg}, the spherical average of any field can be thought of as two angular averages followed by a radial average. Here we compile the expressions for the first of the angular averages, namely over the azimuthal angle $\phi$, for the tidal tensor \eqref{eq:TTcoordinvar}.

Focusing on a single but generic harmonic coefficient $\psi_{\ell m}(r)$ and using \eqn{eq:Tpol-to-Txyz} leads to 
\begin{align}
\avg{T_{xx}}_\phi &= \frac12\delta_{m,0}\left(\tilde A_+ + \tilde T_{\phi\phi}\right) +\frac14\left(\delta_{m,2}+\delta_{m,-2}\right)\left(\tilde A_+ - \tilde T_{\phi\phi}\right) \notag\\
&\ph{\frac12\delta_{m,0}\left(\tilde A_+ + \tilde T_{\phi\phi}\right)}
- \frac{i}2\left(\delta_{m,2}-\delta_{m,-2}\right)\tilde B_+\notag\\
\avg{T_{yy}}_\phi &= \frac12\delta_{m,0}\left(\tilde A_+ + \tilde T_{\phi\phi}\right) -\frac14\left(\delta_{m,2}+\delta_{m,-2}\right)\left(\tilde A_+ - \tilde T_{\phi\phi}\right) \notag\\ 
&\ph{\frac12\delta_{m,0}\left(\tilde A_+ + \tilde T_{\phi\phi}\right)}
+ \frac{i}2\left(\delta_{m,2}-\delta_{m,-2}\right)\tilde B_+\notag\\
\avg{T_{zz}}_\phi &= \delta_{m,0}\,\tilde A_-\notag\\
\avg{T_{xy}}_\phi &= \frac{i}4\left(\delta_{m,2}-\delta_{m,-2}\right)\left(\tilde A_+-\tilde T_{\phi\phi}\right) + \frac12\left(\delta_{m,2}+\delta_{m,-2}\right)\tilde B_+\notag\\
\avg{T_{xz}}_\phi &= \frac12\left(\delta_{m,1}+\delta_{m,-1}\right)\tilde C - \frac{i}2\left(\delta_{m,1}-\delta_{m,-1}\right)\tilde B_-\notag\\
\avg{T_{yz}}_\phi &= \frac{i}2\left(\delta_{m,1}-\delta_{m,-1}\right)\tilde C - \frac12\left(\delta_{m,1}+\delta_{m,-1}\right)\tilde B_-\,,
\label{eq:Txyz-phiavg}
\end{align}
where $\tilde A_{\pm},\tilde B_{\pm},\tilde C$ and $\tilde T_{\phi\phi}$ are obtained by setting $\phi=0$ in $A_{\pm},B_{\pm},C$ (equation~\ref{eq:A+-B+-C}) and $T_{\phi\phi}$ (equation~\ref{eq:TTpol}), respectively, and we used the identities
\begin{align}
\avg{c_{m^\prime\phi}\,\e{im\phi}}_\phi &= \frac12\left(\delta_{m,m^\prime}+\delta_{m,-m^\prime}\right)\notag\\
\avg{s_{m^\prime\phi}\,\e{im\phi}}_\phi &= \frac{i}2\left(\delta_{m,m^\prime}-\delta_{m,-m^\prime}\right)\,,
\label{eq:phiavgs}
\end{align}
where $m^\prime\geq0$. 

\subsection{Axisymmetric model}
\label{app:axisymm}
\noindent
For the reasons discussed in the main text, we focus attention on the axisymmetric case in which all multipoles with $m\neq0$ vanish. This leads to
\begin{align}
\avg{T_{xx}}_\Omega &= \frac12\avg{\left(\tilde A_++\tilde T_{\phi\phi}\right)}_\mu = \avg{T_{yy}}_\Omega\notag\\
\avg{T_{zz}}_\Omega &= \avg{\tilde A_-}_\mu\,,
\label{eq:<Tij>_Omega:ell0}
\end{align}
and the remaining angle-averaged components vanish, so that the angle-averaged tidal tensor is diagonal. 

Using the identities in \eqn{eq:thetaderivs} and suppressing the explicit dependence of $\psi_{\ell0}$ and $P_\ell$ on $r$ and $\mu$, respectively, we find
\begin{align}
&N_{\ell0}^{-1}\,\tilde A_- \notag\\
&= \psi_{\ell0}^{\prime\prime}\,\mu^2P_\ell + \frac1r\,\psi_{\ell0}^\prime\left[P_\ell-(2\ell+1)\mu^2P_\ell+2\ell\,\mu\, P_{\ell-1}\right]\notag\\
&\ph{\psi_{\ell0}^{\prime\prime}}
+\frac1{r^2}\,\psi_{\ell0}\left[-\ell(\ell+1)P_\ell + \ell(\ell+2)\mu^2P_\ell - \ell\,\mu\,P_{\ell-1}\right]\notag\\
&N_{\ell0}^{-1}\,\left(\tilde A_++\tilde T_{\phi\phi}\right) \notag\\
&= \psi_{\ell0}^{\prime\prime}\left(1-\mu^2\right)P_\ell + \frac1r\,\psi_{\ell0}^\prime\left[P_\ell+(2\ell+1)\mu^2P_\ell-2\ell\,\mu\, P_{\ell-1}\right]\notag\\
&\ph{\psi_{\ell0}^{\prime\prime}}
+\frac1{r^2}\,\psi_{\ell0}\left[ - \ell(\ell+2)\mu^2P_\ell + \ell\,\mu\,P_{\ell-1}\right]
\label{eq:A+-Tphiphi:ell0}
\end{align}
where a prime denotes a derivative with respect to $r$. 
As a check, note that 
\begin{align}
\tilde A_-+\tilde A_++\tilde T_{\phi\phi} &= N_{\ell0}P_\ell(\mu)\left[\psi_{\ell0}^{\prime\prime}+\frac2r\,\psi_{\ell0}^\prime-\frac{1}{r^2}\,\ell(\ell+1)\psi_{\ell0}\right] \notag\\
&= \avg{\nabla^2\left(N_{\ell0}\psi_{\ell0}P_\ell\right)}_\phi = \avg{\nabla^2\psi(\mb{r})}_\phi\,,
\label{eq:tracecheck:ell0}
\end{align}
which correctly recovers the azimuthally averaged trace of the tidal tensor.

To average \eqn{eq:A+-Tphiphi:ell0} over $\mu$, we note that $P_1(\mu)=\mu$ and $P_2(\mu)=(3\mu^2-1)/2$, so that \eqns{eq:A+-Tphiphi:ell0} can be rewritten as \begin{align}
&N_{\ell0}^{-1}\,\tilde A_- \notag\\
&= \frac13\,\psi_{\ell0}^{\prime\prime}\,P_\ell\left(2P_2+1\right) \notag\\
&\ph{\psi_{\ell}}
+\frac1r\,\psi_{\ell0}^\prime\left[P_\ell-\frac13\,(2\ell+1)P_\ell\left(2P_2+1\right)+2\ell\,P_1\, P_{\ell-1}\right]\notag\\
&\ph{\psi_{\ell0}^{\prime\prime}}
+\frac1{r^2}\,\psi_{\ell0}\bigg[-\ell(\ell+1)P_\ell + \frac13\,\ell(\ell+2)P_\ell\left(2P_2+1\right) \notag\\
&\ph{\psi_{\ell0}^{\prime\prime}+\frac1{r^2}\,\psi_{\ell0}-\ell+1}
- \ell\,P_1\,P_{\ell-1}\bigg]\notag\\
&N_{\ell0}^{-1}\,\left(\tilde A_++\tilde T_{\phi\phi}\right) \notag\\
&= -\frac23\,\psi_{\ell0}^{\prime\prime}\,P_\ell\left(P_2-1\right) \notag\\
&\ph{\psi}
+\frac1r\,\psi_{\ell0}^\prime\left[P_\ell+\frac13\,(2\ell+1)P_\ell\left(2P_2+1\right)-2\ell\,P_1\, P_{\ell-1}\right]\notag\\
&\ph{\psi_{\ell}}
+\frac1{r^2}\,\psi_{\ell0}\left[ - \frac13\,\ell(\ell+2)P_\ell\left(2P_2+1\right) + \ell\,P_1\,P_{\ell-1}\right]\,.
\label{eq:A+-Tphiphi:ell0-alt}
\end{align}
Using the relations~\eqref{eq:Legendreint} and~\eqref{eq:Legendreortho} then kills all multipoles except $\ell=0$ and $\ell=2$, and we have
\begin{align}
\avg{\tilde A_-}_\mu &= \frac{N_{00}}3\,\delta_{\ell,0}\left(\psi_{00}^{\prime\prime}+\frac2r\,\psi_{00}^\prime\right) \notag\\
&\ph{\frac{N_{00}}3}
+\frac{2N_{20}}{15}\,\delta_{\ell,2}\left(\psi_{20}^{\prime\prime}+\frac5r\,\psi_{20}^\prime+\frac3r\,\psi_{20}\right)\notag\\
\avg{\tilde A_++\tilde T_{\phi\phi}}_\mu &= \frac{2N_{00}}3\,\delta_{\ell,0}\left(\psi_{00}^{\prime\prime}+\frac2r\,\psi_{00}^\prime\right) \notag\\
&\ph{N_{00}}
-\frac{2N_{20}}{15}\,\delta_{\ell,2}\left(\psi_{20}^{\prime\prime}+\frac5r\,\psi_{20}^\prime+\frac3r\,\psi_{20}\right)\,.
\label{eq:A+-Tphiphi:ell0:muavg}
\end{align}
Recognising $\avg{\Delta}_\Omega(r)-1$ as given in \eqn{eq:<Delta>_Omega} in the monopole term in \eqns{eq:A+-Tphiphi:ell0:muavg} and defining $t(r)$ using \eqn{eq:tdef} then leads to \eqns{eq:<Tij>_Omega:ell0-explicit}.

Departing from $m=0$ would mean accounting for $m=\pm1,\pm2$ terms in \eqns{eq:Txyz-phiavg}, whose averages over $\mu$ and $r$ would not only modify the diagonal terms in \eqn{eq:<Tij>(<R):axisymm}, but also introduce off-diagonal terms in general. The calculation of $q^2(<R)$ and hence $\alpha(<R)$ would then need to use \eqn{eq:I1I2} in its full glory.

\section{Setting up the Gaussian mixture}
\label{app:GMdetails}
\noindent
In this Appendix, we describe our choices for the number of components $N_{\rm c}$ and the widths $\sigma_j$ of each component ($1\leq j\leq N_{\rm c}$) for the Gaussian mixture approximation \eqref{eq:NFWgaussmix} to the NFW profile. In principle, one should optimise these quantities so as to produce the smallest possible error in Figure~\ref{fig:error-gmNFW}. In practice, we have found it easier to fix these numbers using the arguments described below, with a \emph{post hoc} justification provided by the accuracy achieved in Figure~\ref{fig:error-gmNFW}. Our subsequent fits are performed using the non-negative least squares (NNLS) algorithm \citep{lh95} implemented in SciPy as {\tt scipy.optimize.nnls}.

The guiding principle we adopt in fitting the NFW profile \eqref{eq:NFWprofile} is to sample the profile at values $x_j$ of the dimensionless variable $x\equiv r/r_{\rm s}$ where the logarithmic slope of the NFW profile matches the corresponding slope of a single Gaussian component of width $\sigma_j$. It is easy to show that this happens when
\be
\sigma_j^2 = \frac{x_j^2(1+x_j)}{(1+3x_j)}\,.
\label{appeq:widthconstraint}
\ee
We start by choosing the minimum and maximum values of $x$, respectively $x_{\rm min}$ and $x_{\rm max}$. The maximum is conservatively set to $x_{\rm max}=10\,c_{\rm vir}$. The minimum is also chosen to scale approximately linearly with concentration, with a mild additional dependence on $R_{\rm vir}$ and $r_{\rm force}$ (set by trial and error to ensure stability over a wide dynamic range) and a hard lower limit at $0.2\,r_{\rm force}/r_{\rm s}$,.

We then choose $N_{\rm c}=2n_{\rm c}+3$ components, with the integer $n_{\rm c}$ to be determined, such that three of the values $\{x_j\}$ are fixed as $x_1=x_{\rm min}$, $x_{2n_{\rm c}+3}=x_{\rm max}$ and $x_{n_{\rm c}+2}=1$. The remaining values then fan outwards symmetrically in $n_{\rm c}$ logarithmic intervals on either side from $x=1$, giving a total of $2n_{\rm c}+3$ components. The logarithmic spacing and number of these intermediate components is chosen so as to give the maximum spread in samples while ensuring that the NNLS algorithm (implemented as below) converges. After some trial and error, for the number of components we settled on using $n_{\rm c}\simeq9$, with a weak dependence on $R_{\rm vir}$ and $r_{\rm force}$ and an upper limit of $n_{\rm c}\leq10$. We also choose to mildly cluster the components close to $x=1$ where the NFW logarithmic slope is changing most rapidly.

The last piece to put in place is the summation constraint \eqref{eq:wkconstraint} for the weights, which we do as follows. Since there are $N_{\rm c}$ weights $\{w_j\}$ to be determined, the standard inversion problem (i.e., without imposing non-negativity of the weight) would be over-determined if we use all $N_{\rm c}$ of the NFW samples as well as the summation constraint. In order to keep the problem well-determined, we therefore start with $N_{\rm c}-1$ of the samples at $x=x_\beta$, $2\leq \beta\leq N_{\rm c}$ as the $(N_{\rm c}-1)\times1$ column vector $V_\beta$, so that
\be
V_\beta = \Delta_{\rm NFW}(r_{\rm s}\,x_\beta|R_{\rm vir},c_{\rm vir})\,;\quad 2\leq \beta\leq N_{\rm c}\,.
\label{appeq:Vdef}
\ee
We then define the $(N_{\rm c}-1)\times N_{\rm c}$ matrix $F$ using
\be
F_{\beta j} = \Delta_{\rm G}(r_{\rm s}\,x_\beta|\sigma_j,r_{\rm s})\,,
\label{appeq:Fdef}
\ee
where $2\leq\beta\leq N_{\rm c}$, $1\leq j\leq N_{\rm c}$ and $\Delta_{\rm G}(r|\sigma_j,r_{\rm s})$ was defined in \eqn{eq:singlegauss}, so that the corresponding $(N_{\rm c}-1)$-dimensional subspace of the Gaussian mixture problem \eqref{eq:NFWgaussmix} can be written as the equation 
\be
F_{\beta j}\,W_j = V_\beta\,.
\notag
\ee
In the absence of the summation constraint, this under-determined problem can be solved using NNLS by minimising the Euclidean norm
\be
\lVert F^{\rm T}F\, W - F^{\rm T}V \rVert
\notag
\ee
leading to an unconstrained least squares estimate of the full $N_{\rm c}$-dimensional vector $W$. This is, of course, not what we want, since the summation constraint is not implemented and the solution would also have highly degenerate component weights due to the under-determined nature of the standard inversion problem.

Written like this, however, the summation constraint is now straightforward to include. We simply augment the vector $F^{\rm T}V$ with the value $1$, and increase the dimension of the matrix $F^{\rm T}F$ by one, adding a row and column with all elements except the last being unity. 
The Gaussian mixture \eqn{eq:NFWgaussmix} augmented by the summation constraint can then be summarised as minimising the norm
\be
\lVert G\, \tilde W - S \rVert\,,
\label{appeq:GM-nnlsnorm}
\ee
where the $(N_{\rm c}+1)\times(N_{\rm c}+1)$ matrix $G$ is given by
\be
G = \left(\begin{array}{ccc|c}
& &&\vdots\\ 
& F^{\rm T} F &&1\\ 
&&&\vdots\\ 
\hline
\cdots&1 &\cdots&0\\ 
\end{array}\right)\,,
\label{appeq:GMmatrix}
\ee
the $(N_{\rm c}+1)\times1$ vector $S$ is 
\be
S = \left(\begin{array}{c}
\\
F^{\rm T}V\\
\\
\hline
1\\ 
\end{array}\right)\,,
\label{appeq:GMsource}
\ee
and the $(N_{\rm c}+1)\times1$ vector $\tilde W$ is 
\be
\tilde W = \left(\begin{array}{c}
\\
W\\
\\
\hline
\tilde w\\ 
\end{array}\right)\,,
\label{appeq:GMvec}
\ee
with $\tilde w$ being a Lagrange multiplier that enforces the summation constraint.
We have used NNLS on \eqn{appeq:GM-nnlsnorm} to produce the results discussed in the main text.

\label{lastpage}

\end{document}